\documentclass[preprint,12pt]{elsarticle}   
\usepackage{graphicx,color,subfigure}
\usepackage{lineno} 
 
\journal{Astroparticle Physics}
\newcommand{\beq}{\begin{equation}}
\newcommand{\eeq}{\end{equation}}
\newcommand{\ba}{\begin{eqnarray}}
\newcommand{\ea}{\end{eqnarray}}

\def\lsim{\raise0.3ex\hbox{$\;<$\kern-0.75em\raise-1.1ex\hbox{$\sim\;$}}}
\def\gsim{\raise0.3ex\hbox{$\;>$\kern-0.75em\raise-1.1ex\hbox{$\sim\;$}}}

\def\theta{\vartheta}

\begin{document}
\begin{frontmatter}

\title{Extragalactic cosmic rays and their signatures} 

\author{V. Berezinsky}

\address{INFN, National Gran Sasso Laboratory, I-67010, Assergi
(AQ) Italy}
\begin{abstract}
The signatures of UHE proton propagation through CMB radiation are 
pair-production dip and GZK cutoff. The visible manifestations of these 
two spectral features are ankle, which is intrinsic part of the dip, 
beginning of GZK cutoff in the differential spectrum and $E_{1/2}$ in 
integral spectrum. Observed practically in all experiments since 1963, 
the ankle is usually interpreted as a feature caused by transition from 
galactic to extragalactic cosmic rays. Using the mass composition measured 
by HiRes, Telescope Array and Auger detectors at energy (1 - 3)~EeV,
calculated anisotropy of galactic cosmic rays at these energies, and 
the elongation curves we strongly argue against the interpretation of the 
ankle given above. The transition must occur  at lower energy, most 
probably at the second knee as the dip model predicts. The other prediction 
of the dip model, the shape of the dip, is well confirmed by HiRes, 
Telescope Array (TA), AGASA and Yakutsk detectors, and, after recalibration 
of energies, by Auger detector. Predicted beginning of GZK cutoff and 
$E_{1/2}$ agree well with HiRes and TA data. However, directly measured 
mass composition remains a puzzle. While HiRes and TA detectors observe 
the proton-dominated mass composition, as required by the dip model, the 
data of Auger detector strongly evidence for nuclei mass composition 
becoming progressively heavier at energy higher than 4~EeV and reaching 
Iron at energy about 35~EeV. The Auger-based scenario is consistent with 
another interpretation of the ankle at energy $E_a \approx 4$~EeV as 
transition from extragalactic protons to extragalactic nuclei. The heavy 
- nuclei dominance at higher energies may be provided by low-energy of 
acceleration for protons $E_p^{\max} \sim 4 $~EeV and rigidity-dependent 
$E_A^{\max} =Z E_p^{\max}$ for nuclei. The highest energy suppression may
be explained as nuclei-photodisintegration cutoff.

\end{abstract}

\end{frontmatter}
%
%
\section{Historical review: galactic or extragalactic origin?}
\label{introduction}
Do observed cosmic rays (CR) have galactic or extragalactic origin?\\
Since late 1950s this question divided theorists in two groups. 
V.L.~Ginzburg \cite{Ginzburg1953} and S.I.~Syrovatsky\cite{Ginz-Syr1960}, 
defended the Galactic origin, while F.~Hoyle \cite{Hoyle1959}, 
G.R.~Burbidge \cite{Burbidge1962,BH1964} and their collaborators suggested 
the extragalactic origin. In 1963 in book \cite{GSbook} Ginzburg and 
Syrovatsky presented the strong arguments in favour of galactic origin, 
but hot discussions between these two groups continued at all International 
Cosmic Ray Conferences (ICRC) until late 1980s.

The main issues of extragalactic origin have been acceleration,
diffusive propagation of extragalactic particles and magnetic fields
in galaxies and clusters. The main point of extragalactic models was  
proposal of Active Galactic Nuclei (AGN) as the sources, and explanation 
of observed isotropy of cosmic rays. Much attention was given to 
acceleration of protons to highest energies above $10^{19}$~eV,
impossible in Galactic model. An upper limit to the maximum energy 
of acceleration was found in \cite{BH1964} from condition of equality 
of the Larmor radius and dimension of the accelerating site, which
later was used in the famous ``Hillas plot'' \cite{HillasPlot}.   
As a plausible model in \cite{Burbidge1962} was proposed the 
the Local Supercluster with AGN observed there.  

V.L. Ginzburg suggested the rigorous test for extragalactic origin, 
predicting the lower-limit for gamma-ray flux from Small Magellanic 
Cloud (SMC). The gas density in this source is fairly well known, 
while the CR flux in extragalactic model is that measured in Milky 
Way (MW). Therefore, the produced gamma-ray flux at energies 
$E_{\gamma} \gsim 100$~MeV is well predicted. The calculated 
gamma-ray flux from $pp \to \pi^0$ production gives the lower limit, 
because part of the detected flux can be produced by electrons and by 
the point-like sources. In 1993 EGRET \cite{EGRET-SMC} put the upper 
limit to gamma-ray flux from SMC well below the Ginzburg lower limit, 
which means that extragalactic CR flux is lower than that observed in 
MW. In 2008 Fermi/LAT \cite{Fermi-SMC} measured gamma-ray flux from SMC 
with a high accuracy, reliably excluding extragalactic origin of the 
bulk of CRs observed in our galaxy. The  exclusion obtained by EGRET 
and FERMI/LAT is not valid for very high CR energies where transition 
to extragalactic CRs may be expected. 

The Standard Model (SM) for Galactic Cosmic Rays was first put forward 
by Ginzburg and Syrovatskii \cite{GSbook} in 1963 and in the main 
features it was accomplished in 1977 - 1978 by discovery of diffusive shock
acceleration \cite{DSA}. The SM is based on (i) supernova remnants as 
sources, (ii) SNR shock acceleration, and (iii) diffusive propagation 
of CR in the Galactic magnetic fields. 

The book \cite{GSbook} was not a review. It described an original and 
detailed construction of Galactic model of cosmic rays and their 
propagation in magnetic fields. The book included the theoretical
issues of synchrotron radiation relative to astrophysics, and diffusion 
equations for propagation of particles with their analytic solutions. 
In particular, the remarkable Syrovatsky analytic solution of diffusion 
equation for point-like sources \cite{SyrSolut} was included there.
However, most important results were obtained phenomenologically. 
The authors put forward the disc-halo model and estimated the magnetic
fields $B$ there, suggesting equipartition relation 
$\omega_{\rm cr} \sim B^2/2\pi \sim \rho u^2/2$, where the first term 
is energy density of cosmic rays and the last one is turbulent energy
density of plasma. Magnetic field in the disc was estimated as 
$B_d \sim 3\times 10^{-6}$~G. Two versions of halo were suggested and 
analysed: the spherical one with $R_h \sim 10 - 15$~kpc and with flattening
halo with $h \sim 3$~kpc. The set of diffusion equations for protons 
and nuclei was solved analytically and, using the data on anisotropy and mass 
composition, diffusion coefficient for disc was estimated as 
$D \sim 1\times 10^{29}$~cm$^2$/s, close to presently known value.     
The authors recognised the problem of $Li,~Be,~B$ as the secondary 
nuclei and using the solutions of diffusion equations for many 
nuclei and galactic models, determined the thickness traversed by
cosmic rays as $x \sim 10$~g/cm$^2$, also close to the present value. 
Finally, the authors estimated the galaxy luminosity in cosmic rays 
$L_{\rm cr}$ using two formulae which at present can be unified  as 
\beq
L_{\rm cr} \sim c \omega_{\rm cr} M_g/x
\label{eq:Lcr}
\eeq
where $M_g$ is the mass of galactic gas. Eq.~(\ref{eq:Lcr}) gives 
$L_{\rm cr} \sim 2\times 10^{40}$~erg/s in a good agreement with 
energy release of SN in our Galaxy.  
 
The most interesting and dramatic story in development of SM is
connected with acceleration. 

In 1977-1978 four works on acceleration at shocks \cite{DSA} appeared
almost simultaneously. The particles are accelerated due to multiple  
reflections from the shock front and may  acquire maximum energy high
enough for galactic cosmic rays. In 1983 Laggage and Cesarsky 
\cite{LC1983} demonstrated that the time of acceleration cycle increases 
during process of acceleration and $E_{\max}$ remains below the observed 
knee, which is galactic feature. It was really a dramatic moment, when
most reliable and beautiful acceleration mechanism seems to be  
not viable. Revival came back not very soon: in 2001 Bell and Lucek 
\cite{Bell} convincingly confirmed the early proposal of Bell \cite{DSA}
about strong amplification of magnetic field upstream by cosmic rays 
themselves due to streaming instability. A highly turbulent field 
with strength up to $B \sim 10^{-4}$~G is produced and increases 
maximum energy up to needed value $E_{\max} \sim 4 Z$~PeV.  
\section{Spectral features and signatures.}
\label{features}
The observed energy spectrum of Cosmic Rays (CR) has approximately a
power-law behavior for $11$ orders of magnitude in energy with several
features that can be linked with particles propagation and acceleration.

The most prominent spectral feature is the {\em knee} in all-particle spectrum
at energy 3-4~PeV, discovered first at the MSU (Moscow State University)
array in 1958 \cite{MSU-knee}. At the knee the spectrum $E^{-\gamma}$
steepens from $\gamma \approx 2.7$ to $\gamma \approx 3.1$. This knee
is provided by the light elements, protons and Helium, and is
explained in the framework of the Standard Model (SM) for Galactic Cosmic
Rays (GCR) by the maximum energy $E_{\max}$ of acceleration in the Galactic
Sources  (Supernovae Remnants, SNR). In the case of the rigidity-dependent
acceleration  $E_{\max} \propto Z$, where $Z$ is charge number of a
nuclei, the maximum acceleration energy  is reached by Iron nuclei, and
thus the Iron knee is predicted to be located at energy by factor 26 times
higher than for proton knee, i.e. at energy $E_{\max}^{\rm Fe} \sim
(80 - 100)$~PeV. Recently, the Iron knee was found indeed at energy 80~PeV
in KASCADE-Grande experiment \cite{kascadeFe} in a good agreement
with rigidity-acceleration prediction.

Above the knee, at energy $E_{\rm skn} \approx (0.4 - 0.7)$~EeV, there
is a faint feature in the spectrum \cite{second-knee} called the {\em
second knee}. It is seen in many experiments (for a review see
\cite{2knee-ankle-rev}). This feature is often interpreted as a
place of transition from galactic to extragalactic CRs.
\begin{figure}[h]
\begin{center}
\includegraphics[height=8.0cm,width=8cm]{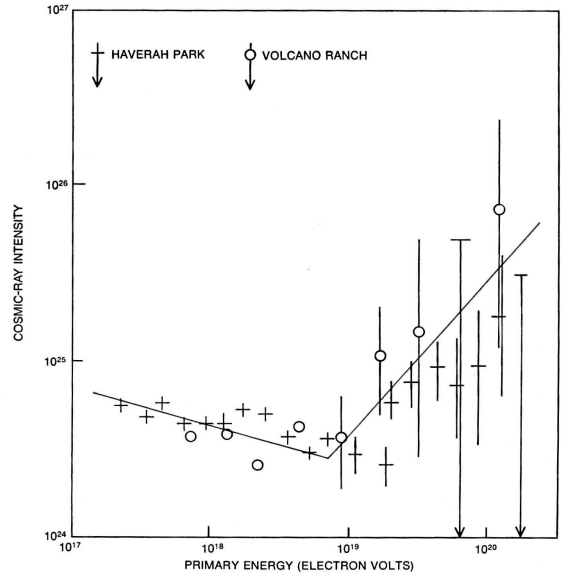}
\caption{The ankle as presented by J. Linsley in review 
\cite{ankle-Linsley} and taken from paper by T.K.~Gaisser \cite{Gaisser}.} 
\label{fig1}
\end{center}
\end{figure}
However, for the last forty years the standard place for transition
from galactic to extragalactic CRs is considered at {\em ankle},
a very prominent spectral feature, observed first
in 1960s by Volcano Ranch detector at energy $E_a^{\rm obs} \sim 
10$~EeV, and it was immediately
interpreted by J. Linsley \cite{Linsley} as transition between
these two components of CRs. This interpretation was further confirmed
by detection a particle with energy about 100~EeV \cite{Linsley-HE}. 
The ankle as presented 
by J.~Linsley in 1973 \cite{ankle-Linsley} is shown in 
Fig.~\ref{fig1} according to data of
Volcano-Ranch and Haverah-Park. (Fig.~\ref{fig1} is taken from paper
by T.K.~Gaisser \cite{Gaisser}). 
At present beginning of ankle is found at 
$E_a^{\rm obs} \approx 4.5 \pm 0.5$~ EeV according to HiRes
observations \cite{ankle-hires}, at $(4.9 \pm 0.3)$~ EeV in  Telescope 
Array  (TA) \cite{ankle-TA}, and at $(4.2 \pm 0.1)$~ EeV in Auger (PAO) 
\cite{ankle-PAO,Kampert}.

What makes us think that CR at highest energies are extragalactic  
and that transition occurs at ankle? 

If CRs at highest energies are galactic heavy nuclei up to Iron, anisotropy 
can be not easily observable. However, SNRs as the sources cannot
provide particles with energies as high as 100~EeV, and for all-galactic-CR 
model one needs additional class of sources which are able to accelerate
particles to much higher energies than SNR.  These sources can be 
hypernovae explosions (GRBs) which occur very rarely, e.g. one per million 
years in our galaxy. Such a model was developed in \cite{Dermer}. More 
recently another galactic model \cite{Kusenko-GRB} also with GRBs as the 
sources was studied, with protons and Iron nuclei as accelerated 
particles and assuming diffusion of Iron nuclei in the galactic magnetic
fields. The calculated energy spectrum explains well the observed PAO 
spectrum. 
  
To prove that observed particles at highest energies are extragalactic, 
one must know their signatures, and these signatures have been found
theoretically in famous works by K.~Greisen \cite{Greisen-gzk} and 
G.T.~Zatsepin and V.A.~Kuzmin \cite{ZK-gzk}, who predicted 
that extragalactic protons propagating through CMB radiation obtain
the specific steepening of energy spectrum, called Greisen-Zatsepin-Kuzmin 
(GZK) cutoff.

The highest energy feature, the steepening ('cutoff') of the
spectrum is found indeed in all three largest detectors, HiRes
\cite{GZK-hires}, Telescope Array (TA) \cite{GZK-TA} and
PAO \cite{GZK-PAO}, though the nature of this cutoff is still
questionable.

\begin{figure}[ht]
\begin{center}
 \begin{minipage}[ht]{60 mm}
 \centering
 \vspace{5mm}
 \includegraphics[width=60 mm,height=80mm]{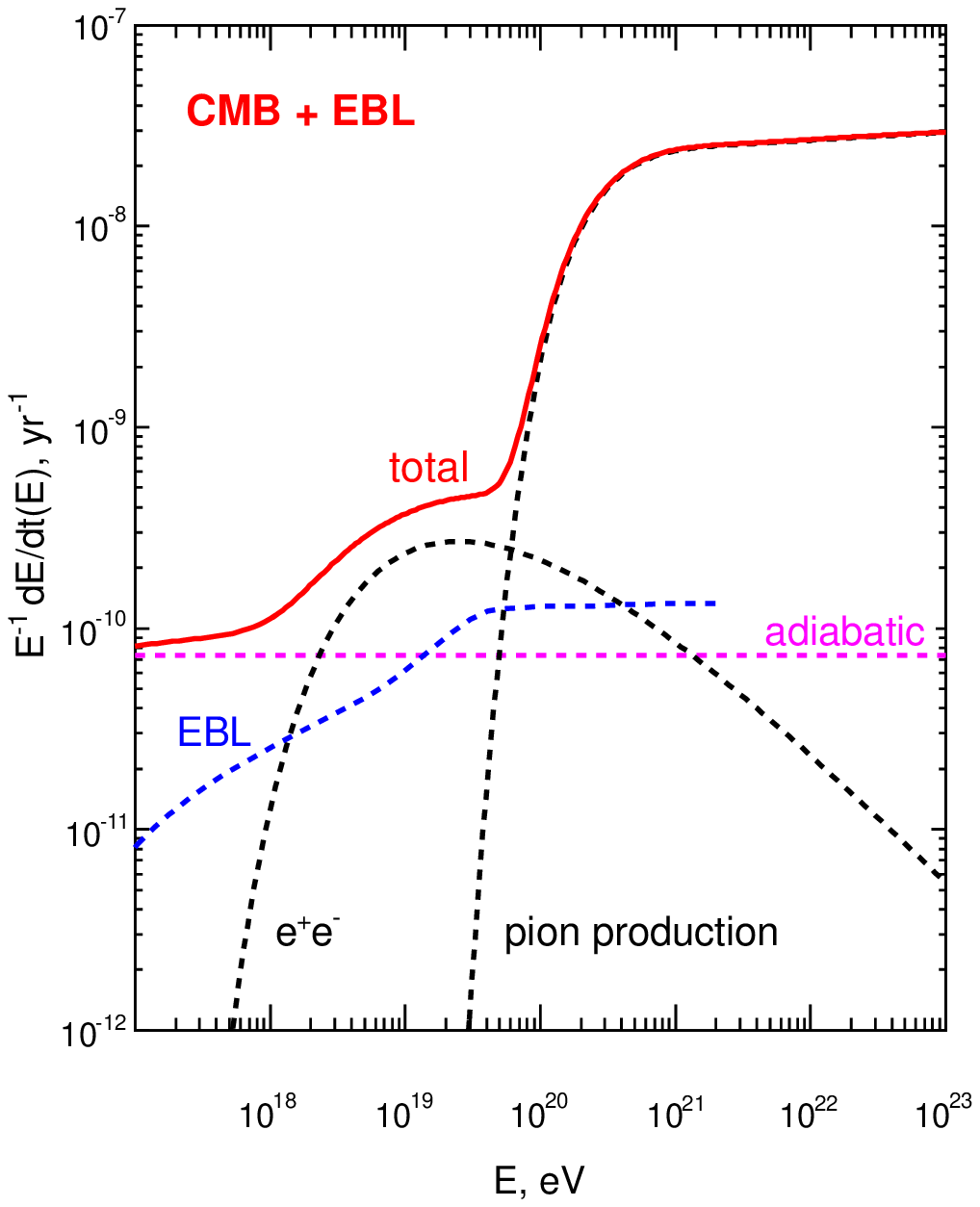}
 \end{minipage}
 \hspace{1mm}
 \vspace{-1mm}
 \begin{minipage}[h]{60 mm}
 \centering
 \includegraphics[width=60 mm]{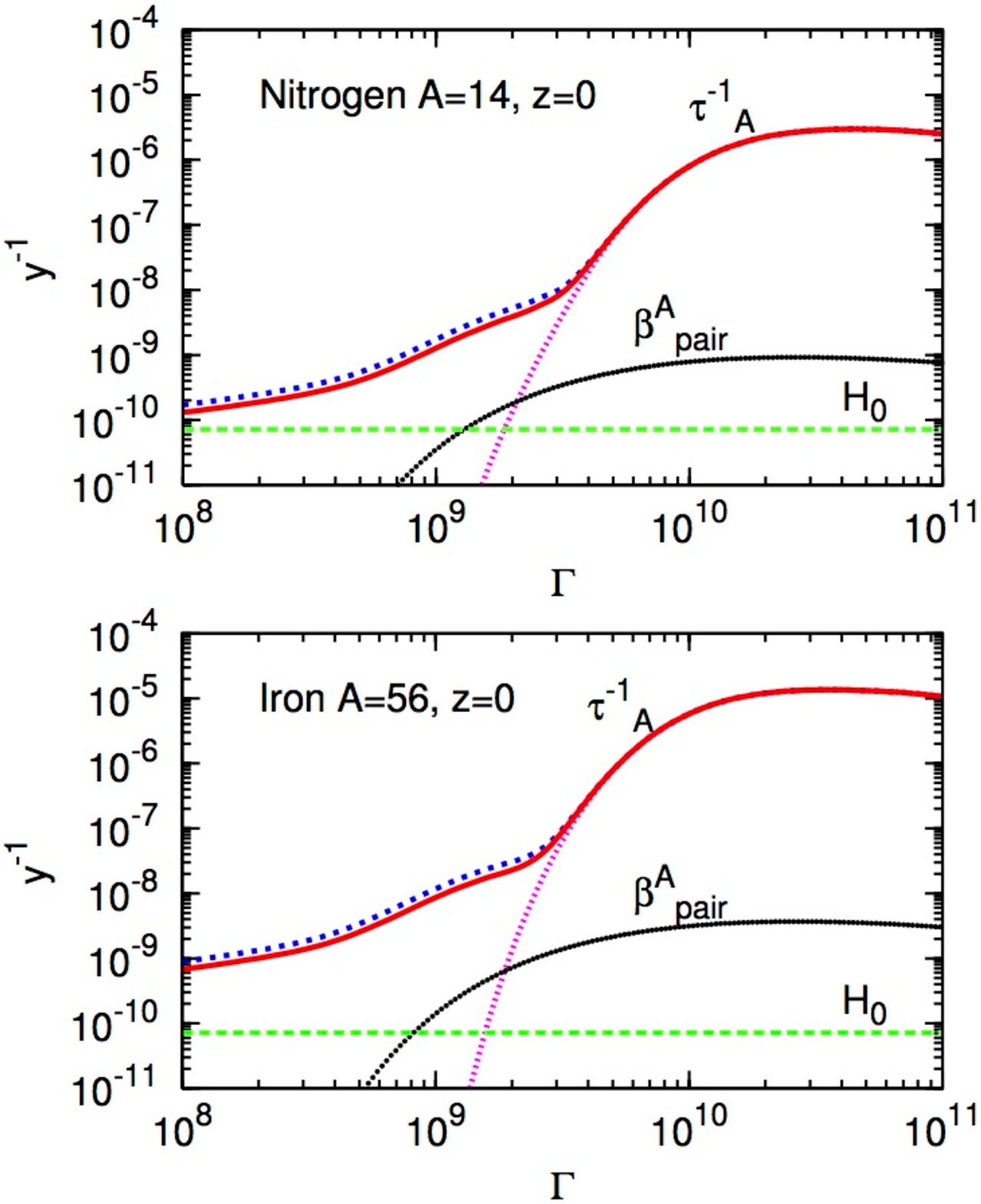}
 \end{minipage}
 \end{center}
\vspace{-4 mm}%
\caption{
Energy losses of protons (left panel) on CMB and EBL. One can see that 
energy losses on EBL can be neglected. There are two characteristic energies: 
$E_{\rm eq1} = 2.4$~EeV (intersection of pair-production $e^+e^-$ 
and adiabatic curves) and $E_{\rm eq2}=61$~EeV (intersection of pair-
and pion- production curves). It is clear that
beginning of the GZK cutoff must be about (but below) $E_{\rm q2}$. 
In the right panel the inverse lifetimes of two nuclei, Nitrogen and 
Iron, are shown as function of Lorentz factor $\Gamma$ from \cite{ABG1,ABG2}. 
The dotted curves show contribution of EBL. The small-dots lines show
the case of CMB only, which corresponds to  steepening  at
intersection of pair-production curve (line $\beta^A_{\rm pair}$) with 
adiabatic curve $H_0$. Presence of EBL strongly modifies this picture. 
See text for more details.
}
\label{fig2}
\end{figure}

The spectrum steepening is quite different for protons and nuclei as 
primaries. The energy losses and lifetimes of these particles interacting 
with CMB and EBL (Extragalactic Background Light) photons are shown in 
Fig.\ref{fig2}. 

The GZK cutoff is caused by photopion production on CMB photons: 
\beq
p + \gamma_{\rm cmb} \rightarrow N + \pi .
\label{eq:gzk}
\eeq
The energy  $E_{\rm eq2} \approx 60$~EeV (see Fig.~\ref{fig2} ) describes 
this steepening as interaction signature. In realistic calculations 
beginning of the GZK cutoff is lower, because at larger $z$  the 
intersection point $E_{\rm eq2}(z)$ is shifted by factor $(1+z)$ to lower 
energies. The spectrum must show steepening at lower energy  
$E_{\rm GZK} \approx 50$~EeV, though this is (weakly) model-dependent 
quantity. However, the PAO data for beginning of steepening 
$\log E_{\rm GZK} = 19.41 \pm 0.02$, with $E$ in eV, (or 
$E_{\rm GZK}=25.7 \pm 1.2$~EeV) given in Table 1 of \cite{GZK-PAO2011} 
is below all known theoretical predictions, even after taking into 
account a possible $22\%$ energy shift due to systematic energy errors.

On the other hand HiRes \cite{GZK-hires} and TA \cite{GZK-TA} data
agree well with predicted beginning of the GZK cutoff and its energy
shape. This contradiction reflects the different mass composition 
found in these experiments. In the discussed energy region the mass
composition of PAO is presented by heavy nuclei and the observed spectrum 
cutoff might be explained as {\em nuclei-photodisintegration} cutoff 
(see nuclei lifetimes in the right panel of Fig.\ref{fig2}). In this
panel the continuous pair-production energy losses $\beta^A_{\rm pair}$ 
and adiabatic energy losses (curve $H_0$) are shown together with lifetime
of photodisintegration on CMB (small-dots lines) and EBL (dotted curves). 
The calculated lifetimes (from recent papers \cite{ABG1} and \cite{ABG2})
correspond to emission of one and more nucleons. The Lorentz-factor 
$\Gamma$ as energy variable in Fig.~\ref{fig2} is chosen because it is
practically not changed in the process of photodisintegration. 
It changes mainly due to pair-production on CMB at large $\Gamma$ and 
adiabatic energy-loss at lower $\Gamma$. If one is interested  in 
any nucleus $A$ produced at propagation of accelerated nucleus $A_0$, 
its energy $E=Am_N \Gamma$  changes in all three processes described 
above (for more details see \cite{ABG1,ABG2} and review \cite{Allard-rev}).   
If to include in consideration only CMB, then $E^{-1}dE/dt$  strongly 
increases after crossing the adiabatic energy loss by pair-production 
curve. Qualitatively one can see it in the right panels of 
Fig.~\ref{fig2} in case of small-dots curves. Then situation reminds 
the GZK cutoff and photodisintegration spectrum-steepening is only a
little lower than in case of GZK. This was first demonstrated in  works 
\cite{BZ1971,BGZ} in 1971 and 1975, respectively.  Presence of EBL  
radically changes this picture, because nuclei are photodisintegrated 
faster than they lose kinetic energy. The dependence of
photodisintegration time on Lorentz-factor becomes smoother (see 
the dotted curves in Fig.~\ref{fig2}). A possible efficient 
photodisintegration of UHE nuclei on extragalactic light was first
demonstrated by N.M.~Gerasimova and I.L.~Rozental  in 1961 \cite{GR}.
     
If one considers the energy spectra with fixed $A$, e.g. the primary
nuclei only, the steepening of the spectrum is produced by increasing
of $\tau_A^{-1}$ with Lorentz-factor which is less sharp in case 
of CMB+EBL than for CMB only (see Fig.~\ref{fig2}). 

The photodisintegration of nuclei on CMB was only shortly
mentioned by K.~Greisen \cite{Greisen-gzk}, and G.T.~Zatsepin  
and V.A.~Kuzmin \cite{ZK-gzk}, the first numerical calculations 
have been performed by F.W.Stecker \cite{Stecker-first} in 1969.
Later Stecker and his collaborators developed this study of nuclei 
propagation through background radiation, providing the basic elements
needed for calculations, namely, the nuclei photodisintegration and 
EBL models with its cosmological evolution \cite{Stecker-nucl}.  
Many practically important calculations of nuclei spectra have been
performed by Allard, Olinto and Parizot, see review \cite{Allard-rev}
for references. 

Historically the GZK cutoff is related to production of pions on CMB 
$p+ \gamma_{\rm cmb} \to N + \pi$. To rename it including 
there also nuclei photodisintegration would be incorrect. Greisen, 
Zatsepin and Kuzmin just mentioned this process, considering it as 
nuclei interaction with CMB photons only. In this case 
beginning of the cutoff and its shape is completely different from the 
case of CMB+EBL, as one can see it from from right panel of Fig.\ref{fig2}.
Authors of GZK cutoff explicitly emphasized connection of their work
with CMB only. The extragalactic light was first considered in \cite{GR}.

In contrast to GZK cutoff the nuclei photodisintegration cutoff has no well
predicted observational signatures: in particular the energy position
and shape of spectrum steepening are different for different nuclei. 
\section{Signatures of proton propagation through CMB.}
\label{signatures}
Propagating through CMB, UHE protons undergo two interactions:
photo-pion production $p+\gamma_{\rm cmb}\to N+ \pi$ and
$e^+e^-$ production $p+\gamma_{\rm cmb}\to p+e^++e^-$.  As a result,
the proton spectrum is distorted: due to the first interaction
it obtains a sharp steepening called {\em GZK cutoff} and due to the
second one - shallow deepening, called {\em dip}. Both features depend
not only on interactions but also on model-dependent quantities, e.g.
on modes of propagation (diffusion or rectilinear propagation), on
cosmological evolution of the sources, on source separation etc. These
model-dependent distortions are especially strong for GZK feature.

{\em The main strategy of this presentation as well as of works
\cite{BGGPL,BGGprd,Aletal} is to distinguish the interaction signatures
from the model-dependent ones.}

This is possible to do using the {\em modification factor}
$\eta(E)$ \cite{BG88}, some kind
of theoretical spectrum, in which model-dependent features are suppressed
or absent. For this aim we use the calculations which involve only one
free physical parameter. One naturally expects that the calculated interaction
signature in terms of modification factor cannot have the agreement with
observations with good $\chi^2$, because the observational data include
the model-dependent features described by many parameters, such as
cosmological evolution of the sources, source separation etc. One free
parameter is not enough to describe 4 - 5 different experiments with
about 20 energy bins in each. As the next step we perform the
{\em model-dependent} calculations which necessarily include many free
parameters improving further the agreement. This analysis should
be done in terms of natural spectrum, $J(E)$ or $E^3J(E)$, where the 
model-dependent features are not suppressed.

However, the Nature has been more kind to us than we expected. The
dip, in terms of modification factor with one free physical parameter
(interaction-signature description) gave very good $\chi^2$ agreement
with observations of four experiments \cite{data}: Yakutsk, AGASA,
HiRes and Telescope Array (see section \ref{modfactor-dip}).
The comparison with PAO data has a different story. Comparison in
terms of modification factor with PAO observational data from ICRC
2007 (Merida) had good enough $\chi^2$ (see \cite{TAUP2007}). However,
as general prediction, the agreement must become worse with increasing
statistics, and comparison of modification factor with PAO data 2010 
has confirmed it. The reason is that for the difference in case of
very small error bars the model-dependent effects are responsible. 
Indeed, as demonstrated in section \ref{energy-calibrator} the 22\% 
shift of PAO energy scale and cosmological evolution make the PAO 
spectrum in terms of  $E^3J(E)$  compatible with the pair-production 
dip with high accuracy. However, this interpretation contradicts the 
nuclei mass composition measured by PAO in energy region of the dip.
\subsection{The dip in terms of modification factor}
\label{modfactor-dip} 
In this section the dip will be studied
as a {\em signature} of UHE proton interaction with CMB, using the
modification factor as a tool. The modification factor $\eta(E)$
is defined as the ratio of proton spectrum $J_p(E)$ calculated
with all energy losses to the so-called unmodified spectrum
$J_{\rm unm}(E)$ in which only adiabatic energy losses (red-shift)
are included: 
\beq 
\eta(E)=J_p(E)/J_{\rm unm}(E) . 
\label{unm}
\eeq 
\begin{figure}[t]
\begin{center}
\includegraphics[height=6.0cm]{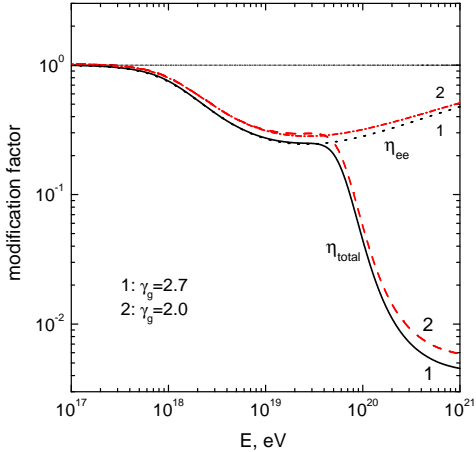} 
\caption{Modification factors for two generation indices $\gamma_g=2.7$
and 2.0. The dotted curve shows $\eta_{\rm ee}$ when only adiabatic and
pair-production energy losses are included. The solid and dashed
curves include also pion-production losses.
}
\label{fig3}
\end{center}
\end{figure}
Modification factor is an excellent tool for {\em interaction  
signature}. As one might see the interactions enter
only numerator and thus they are not suppressed in $\eta(E)$,
while most other phenomena enter both numerator and denominator
and thus they are suppressed or even cancelled in modification
factor, the property being especially pronounced for the dip
modification factor, which according to our calculations
\cite{Aletal} depends very weakly on generation index $\gamma_g$
and $E_{\max}$, on propagation mode, source separation within 1 -
50 Mpc, local source overdensity or deficit etc.  The dip
modification factor is modified strongly by presence of nuclei
($\gsim 15\%$). Modification factor for GZK feature is changing
stronger.

\begin{figure}
\begin{center}
 \begin{minipage}[h]{53mm}
 \includegraphics[width=53mm,height=51mm]{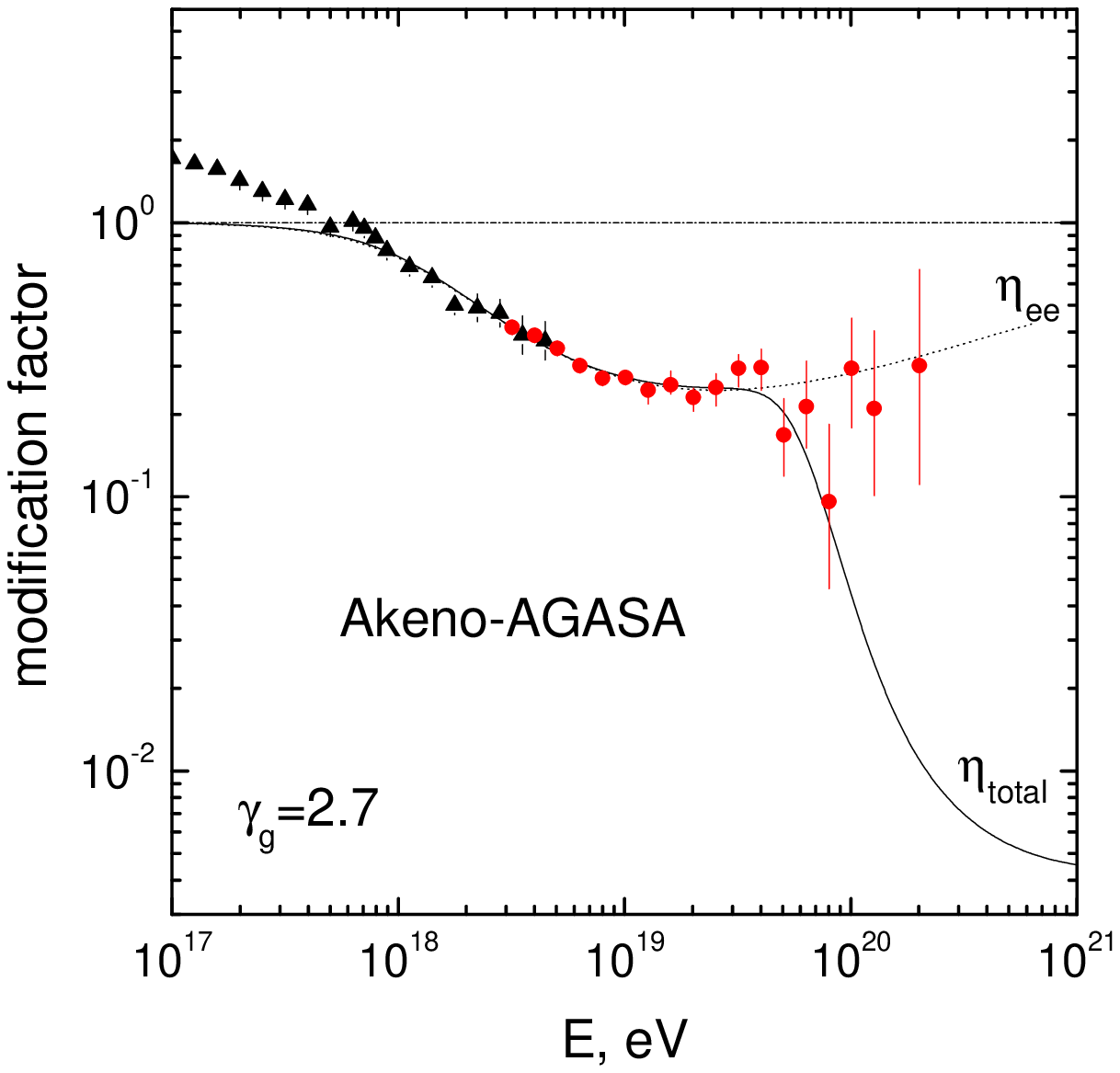}
 \end{minipage}
 \hspace{1mm}
 \begin{minipage}[h]{54 mm}
 \includegraphics[width=54mm,height=51mm]{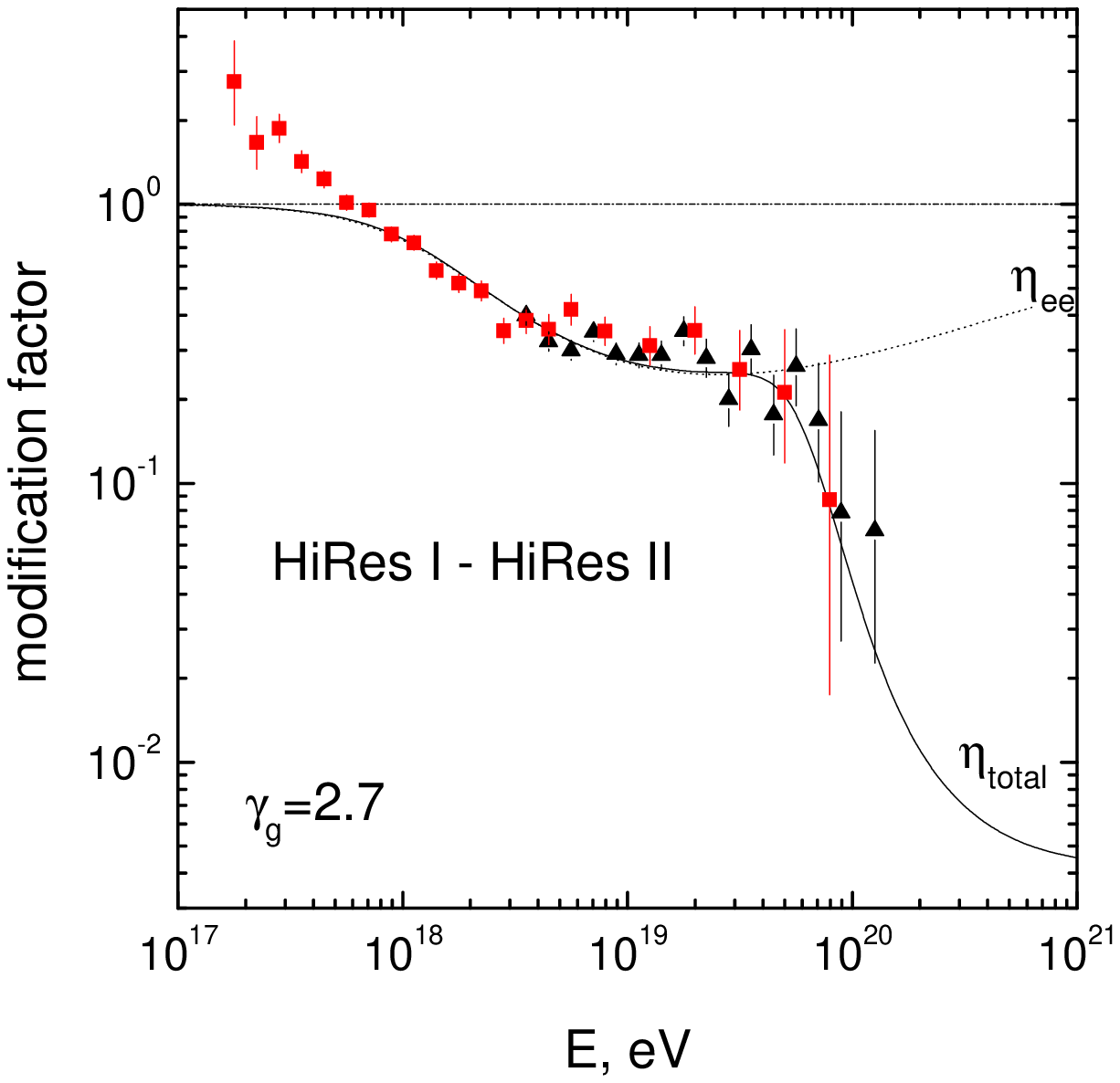}
 \end{minipage}
\newline \noindent
\medskip \hspace{-8mm}
 \begin{minipage}[ht]{54mm}
 \includegraphics[width=53mm,height=53mm]{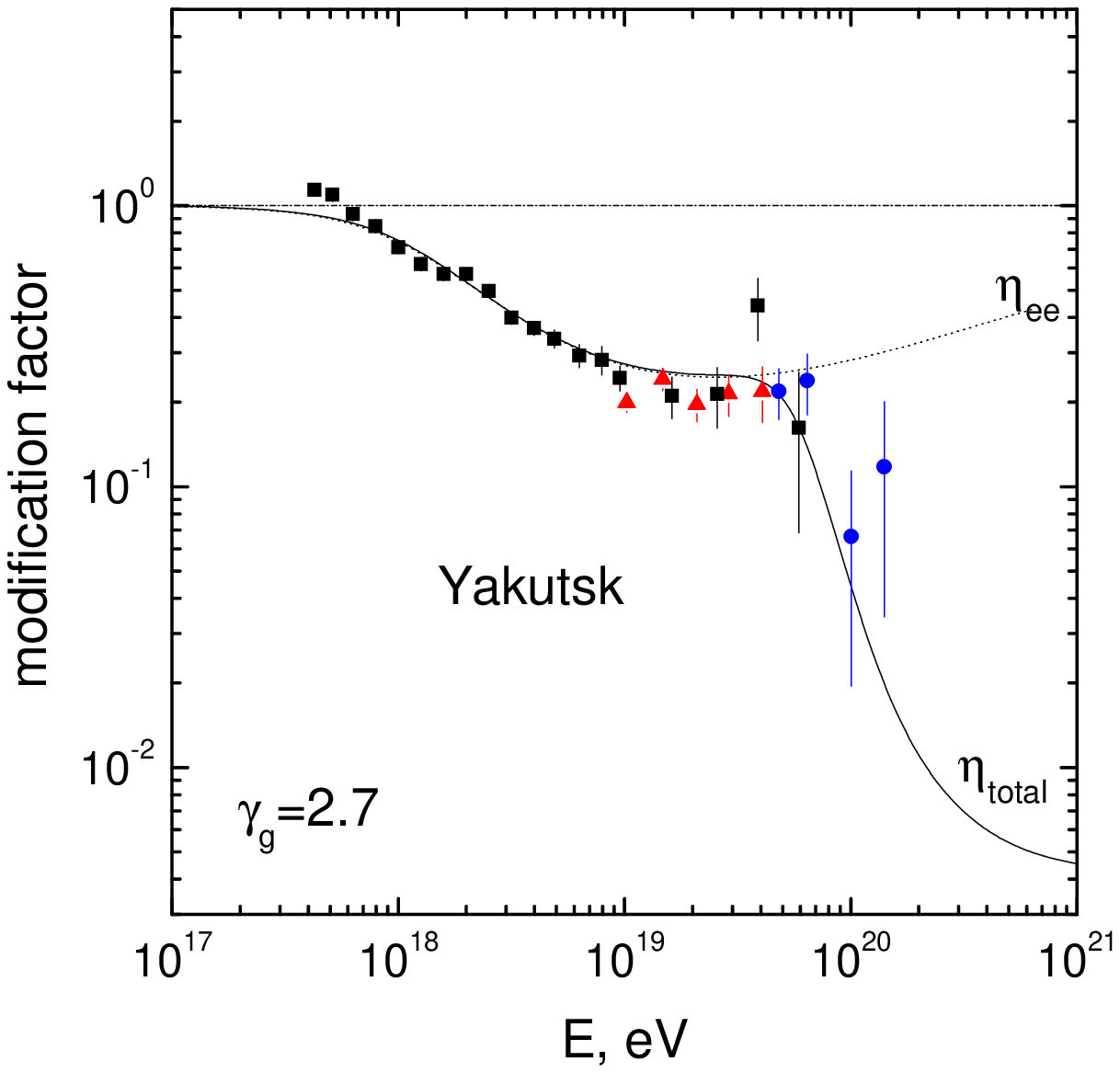}
 \end{minipage}
 \hspace{2mm}
 \begin{minipage}[h]{54 mm}\medskip
 \includegraphics[width=53mm,height=54mm]{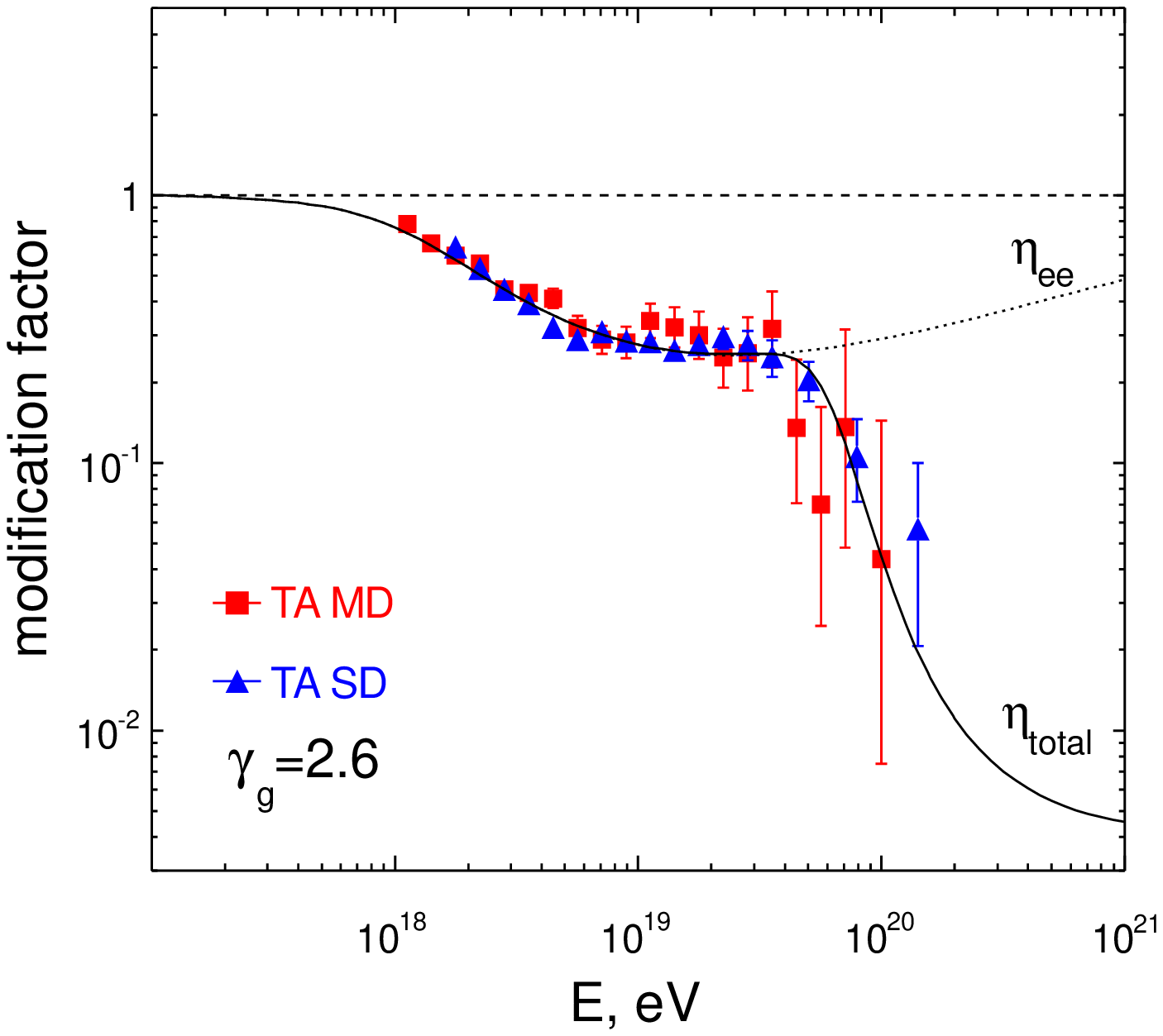}
\end{minipage}
 \vspace{-2 mm}%
\caption{ The predicted pair-production dip in comparison with
Akeno-AGASA, HiRes, Yakutsk and Telescope Array data \cite{data}.
All these experiments confirm the dip behavior with good accuracy,
including also the data of Fly's Eye \cite{data} (not presented
here).
} %
\label{fig4}
\end{center}
\end{figure}

Theoretical modification factors calculated for different source
generation indices $\gamma_g$ are presented in
Fig.~\ref{fig3}. If one includes in the calculation of
$J_p(E)$ only adiabatic energy losses, then, according to its
definition, $\eta(E)=1$ (dash-dot line in
Fig.~\ref{fig3}). When $e^+e^-$-production is
additionally included, one obtains spectrum $\eta(E)$ shown in
Fig.~\ref{fig3} by the curves labeled as $\eta_{ee}$.
With the pion photo-production process being also included, the
GZK feature (curves ``total'') appears. The observable part of the
dip extends from the beginning of the GZK cutoff at $E \approx
40$~EeV down to $E \approx 1$~EeV, where $\eta \approx 1$. It has
two flattenings: one at energy $E_a^{\rm tr} \sim 10$~EeV and the
other at $E_b \sim 1$~EeV. {\em The former automatically produces the
ankle} (see Fig.~\ref{fig4}) and the latter provides an
intersection of the flat extragalactic spectrum at $E \leq 1$~EeV
with the steeper Galactic one.

We discussed above the theoretical modification factor. The
{\em observed} modification factor, according to definition, is given by the
ratio of the observed flux $J_{\rm obs}(E)$ and unmodified spectrum
$J_{\rm unm}(E) \propto E^{-\gamma_g}$, defined up to normalization as:
$\eta_{\rm obs} \propto J_{\rm obs}(E)/E^{-\gamma_g}$.
Here $\gamma_g$ is the exponent of the generation spectrum $Q_{\rm
gen}(E_g) \propto E_g^{-\gamma_g}$ in terms of initial proton energies
$E_g$. Fig.~\ref{fig4} shows that both the pair-production dip and
the beginning of the GZK cutoff up to $80$~EeV are well
confirmed by experimental data \cite{data} of Akeno-AGASA, HiRes,
Yakutsk and Telescope Array (TA).
The comparison of the theoretical dip with observational data includes
only two free parameters: exponent of the power-law generation spectrum
$\gamma_g$ (the best fit corresponds to $\gamma_g=2.6 - 2.7$) and the
normalization constant to fit the $e^+e^-$-production dip to the
measured flux. The number of energy bins in the different experiments is
$20 - 22$. The fit is characterized by $\chi^2/{\rm d.o.f.} = 1.0 - 1.2$
for AGASA, HiRes and Yakutsk data.  This is a very good fit for {\em
interaction signature} (see beginning of this section). For this fit we 
used the modification factor without cosmological evolution of sources. 
As was explained above, using a model approach with additional three
parameters describing the cosmological evolution one can further
improve the agreement.

In Fig.~\ref{fig4} one can see that at $E \lsim 0.6$~EeV the
experimental modification factor, as measured by Akeno and HiRes,
exceeds the theoretical modification factor. Since by definition the
modification factor must be less than one, this excess signals the
appearance of a new component of cosmic rays at $E < E_{\rm tr} \approx
0.6$~EeV, which can be nothing else but the Galactic cosmic rays. This
interpretation is confirmed by transition of heavy component to the
protons in the upper-left panel of Fig.~\ref{fig7}, 
that with good accuracy occurs at the
same energy. Thus, according to HiRes data the transition from
extragalactic to Galactic cosmic rays, occurs at energy $E_{\rm tr}
\sim 0.6$~EeV and is accomplished at $E \sim E_b \approx 1$~EeV (see
upper-left panel in Fig.~\ref{fig7} as example).
\subsection{Pair-production dip as energy calibrator}
\label{energy-calibrator}
The energy position of pair-production dip is rigidly fixed by interaction
with CMB and thus it can serve as energy calibrator for the detectors.

As we already mentioned, it is difficult to expect that in terms of
the modification factor the dip, described by one free physical
parameter, can fit the observational data with minimum $\chi^2$.
One can shift the observed energy bins by the recalibration factor
$\lambda_{\rm cal}$, within the systematic error of observations,
to minimize $\chi^2$ \cite{BGGprd}.
We shall refer to this procedure as 'recalibration of energy scale'.
\begin{figure}[t]\medskip
\begin{center}
 \begin{minipage}[h]{60mm}
 \includegraphics[width=60mm,height=47mm]{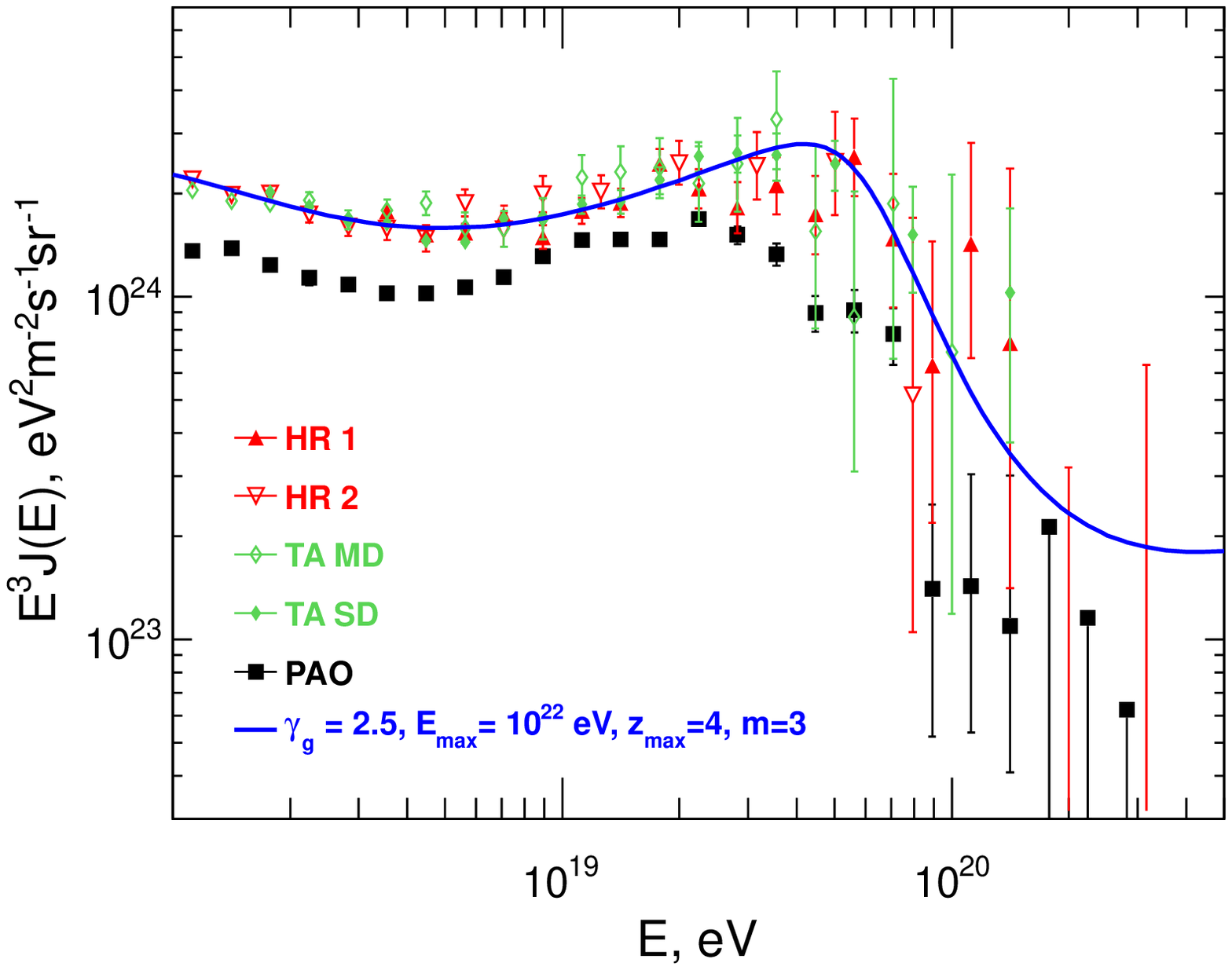}
 \end{minipage}
 \hspace{4mm}
 \begin{minipage}[h]{60mm}
 \includegraphics[width=60mm,height=47mm]{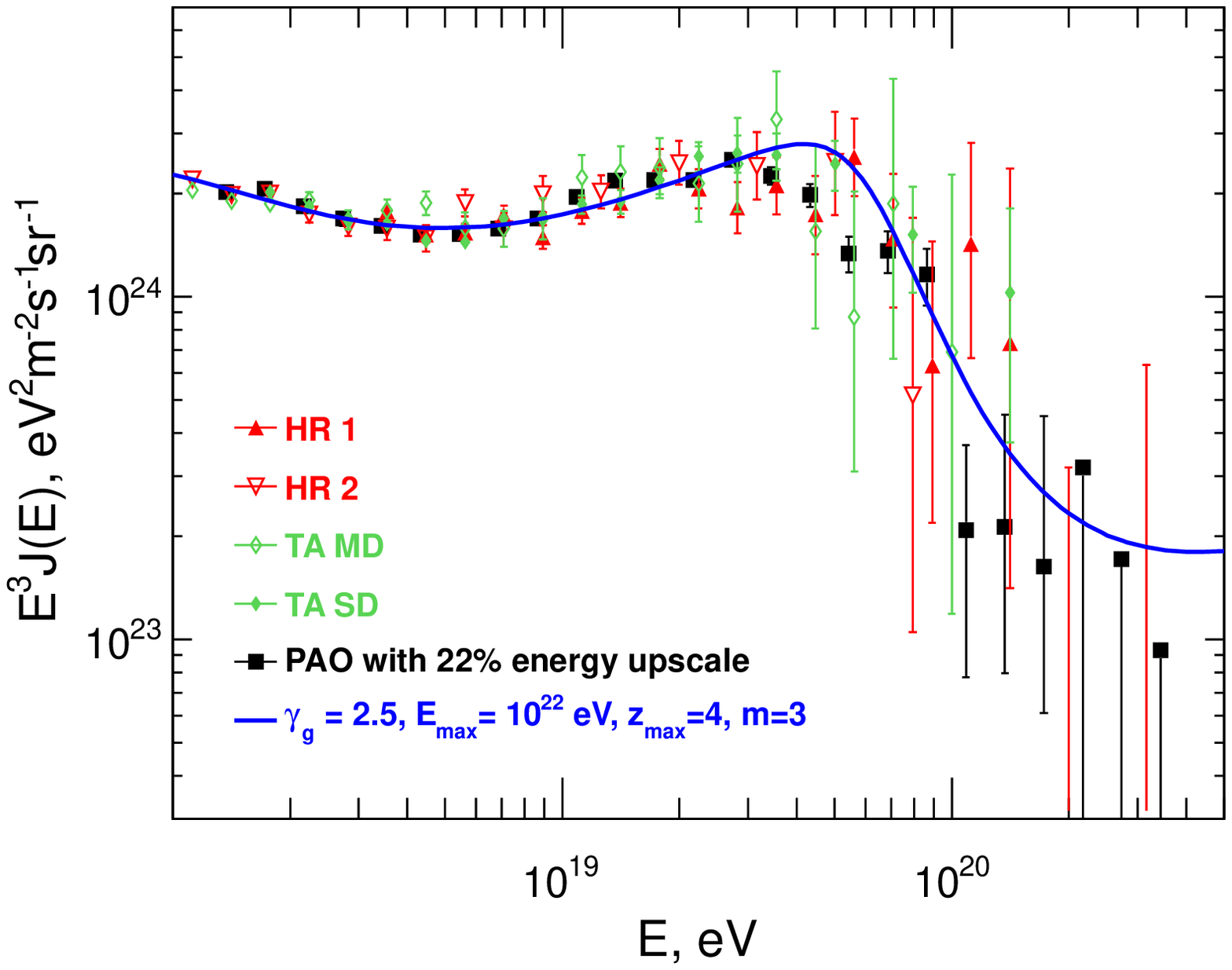}
 \end{minipage}
\caption{{\em Left panel:} Comparison of the PAO energy spectrum
(filled boxes) with the HiRes and TA data fitted by theoretical
pair-production dip (solid curve). {\em Right panel:} Spectra
after energy recalibration of the PAO data with $\lambda_{\rm
cal}=1.22$ (see the text). } 
\label{fig5}
\end{center}
\end{figure}
\begin{figure}[h]\medskip
\begin{center}
 \begin{minipage}[h]{60mm}\hspace{-5mm}
 \includegraphics[width=60mm,height=47mm]{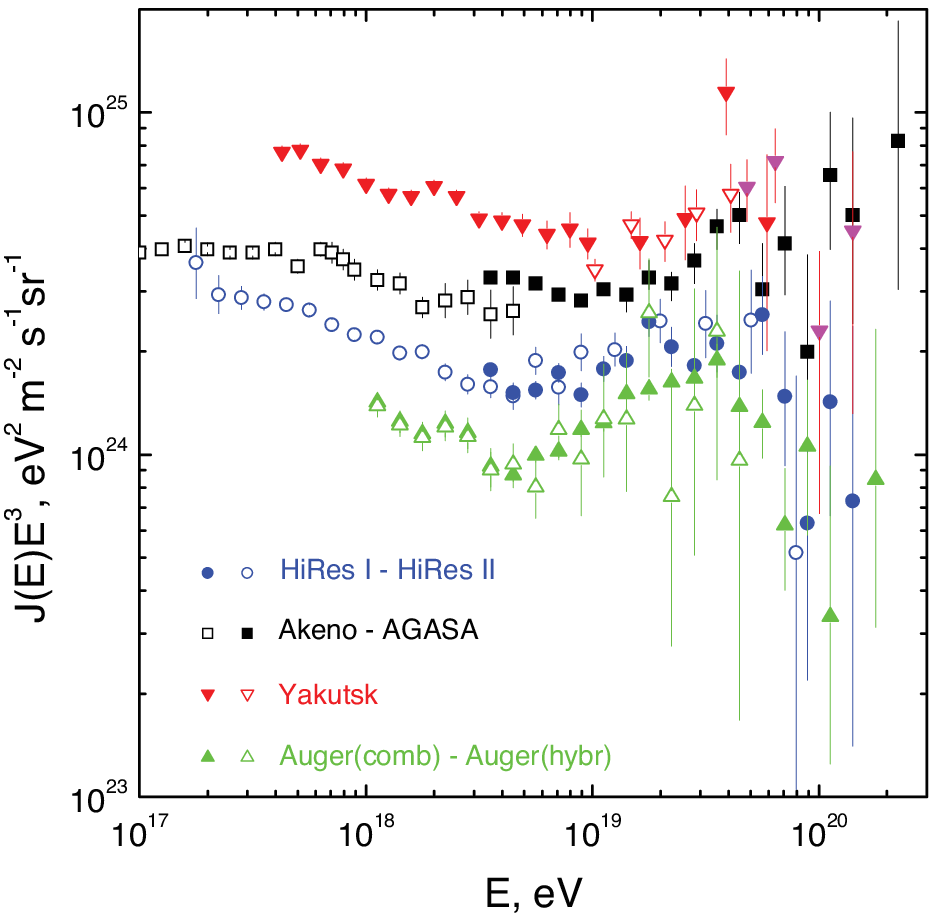}
 \end{minipage}
 \hspace{4mm}
 \begin{minipage}[h]{60mm}\vspace{3mm}\hspace{-5mm}
 \includegraphics[width=60mm,height=45mm]{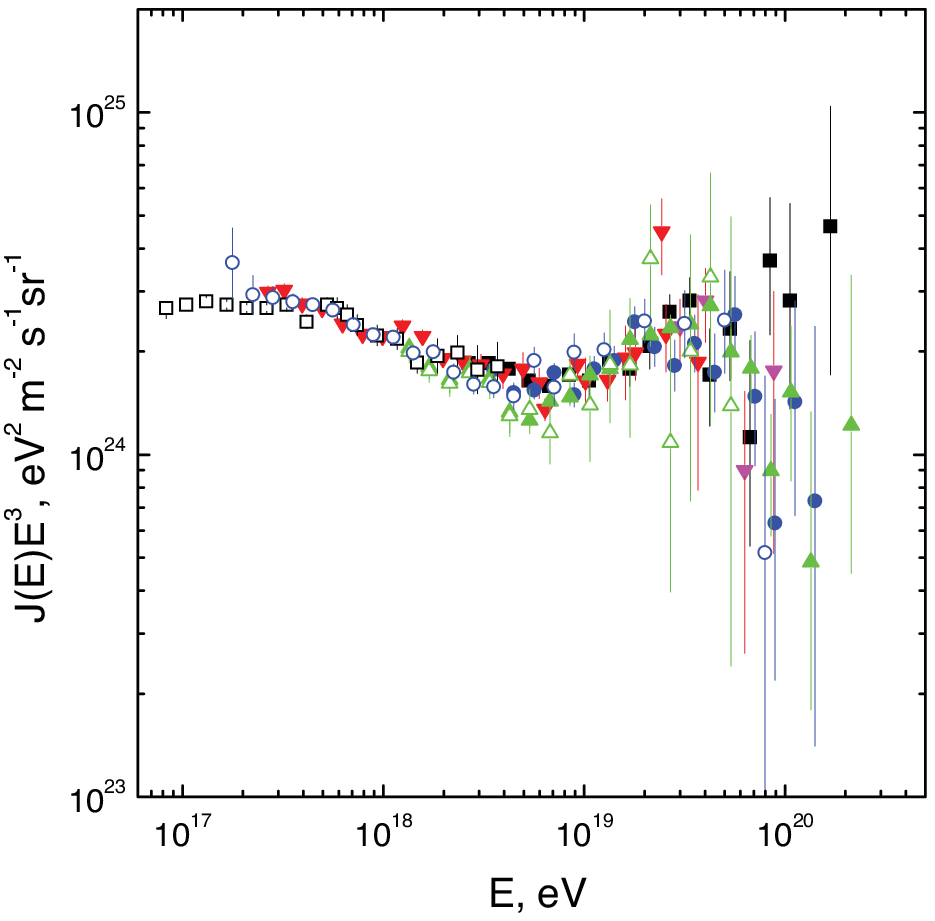}
 \end{minipage}
\caption{{\em Left panel:} Original fluxes from all detectors
(fluxes from HiRes and TA are approximately the same). {\em Right
panel:} Spectra after energy recalibration by pair-production dip:
$\lambda_{\rm cal}=1$ for HiRes/TA, $\lambda_{\rm cal}=1.22$ for
PAO, $\lambda_{\rm cal}=0.75$ for AGASA and $\lambda_{\rm
cal}=0.625$ for Yakutsk. } 
\label{fig6}
\end{center}
\end{figure}
We discuss first the dip in PAO spectrum \cite{GZK-PAO2011}
presented by the filled boxes in the left panel of
Fig.~\ref{fig5}. As it was already mentioned, because of
very small statistical error bars it has too large $\chi^2$ in
comparison with pair-production dip terms of modification factor.
We shall use then the model-dependent method in terms of $E^3J(E)$
including the cosmological evolution $(1+z)^m$ up to $z_{\max}$
as shown in the left panel of Fig.~\ref{fig5}. Using two
more free parameters $m$ and $z_{\max}$ we can reach better agreement
with the modified shape of the dip shown by the solid curve in
Fig.~\ref{fig5}. Now we can shift the PAO energy bins by
factor $\lambda_{\rm cal}$ reaching the minimum $\chi^2$. For this
$\lambda_{\rm cal}=1.22$ is needed. As a result we obtain picture shown in
the right panel of  Fig.~\ref{fig5}. We obtained not only
the excellent agreement with the shape of the theoretical
pair-production dip (solid curve) but also the good agreement with
absolute fluxes of HiRes and TA. Note that disagreement with GZK
cutoff remains for the three energy bins in energy interval $35 - 52$~EeV.  %
Recalibration with help of pair-production dip for all five detectors
(HiRes, Telescope Array, PAO, AGASA and Yakutsk) is shown in
Fig.~\ref{fig6}. Recalibration factor $\lambda_{\rm cal} =1$
for HiRes/TA is based on the scale factor which correctly describes
the pair-production dip and  GZK cutoff in differential and integral
($E_{1/2}$) spectra.
\subsection{GZK cutoff in HiRes and Telescope Array data}
\label{GZK}
The two largest Extensive Air Shower (EAS) detectors, HiRes \cite{GZK-hires}
and Pierre Auger Observatory \cite{GZK-PAO} have observed a sharp
steepening in the UHECR spectrum at $E \gsim (30 - 50)$~EeV. Both
collaborations claimed that the observed steepening is consistent with the
GZK cutoff. But as a matter of fact, there is a dramatic conflict between
these two results, which still leaves the problem open.
\begin{figure}[h]
\begin{center}
 \begin{minipage}[h]{60mm}
 \includegraphics[width=60mm,height=45mm]{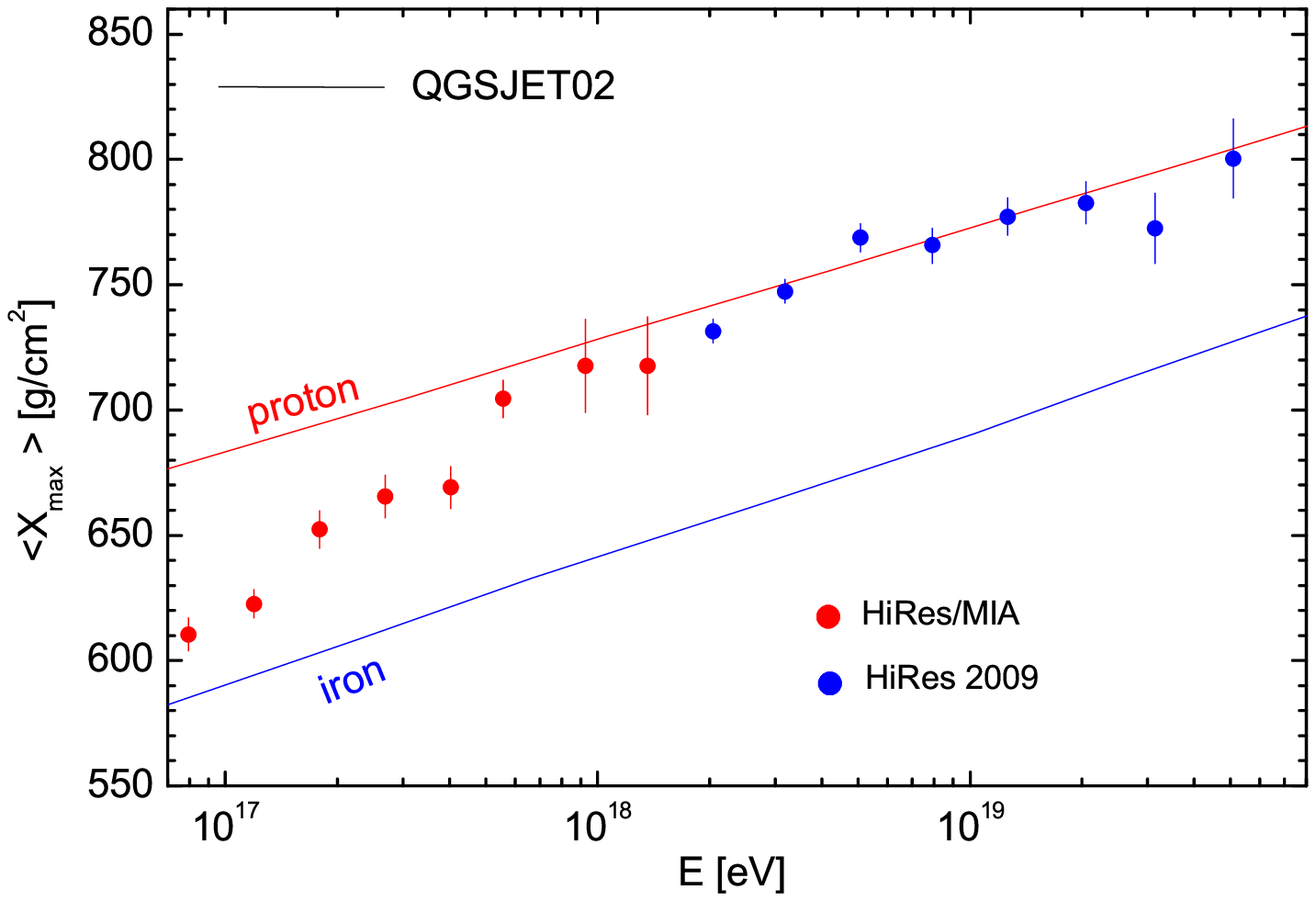}
 \end{minipage}
 \hspace{3mm}
 \begin{minipage}[h]{60mm}
 \includegraphics[width=60mm,height=45mm]{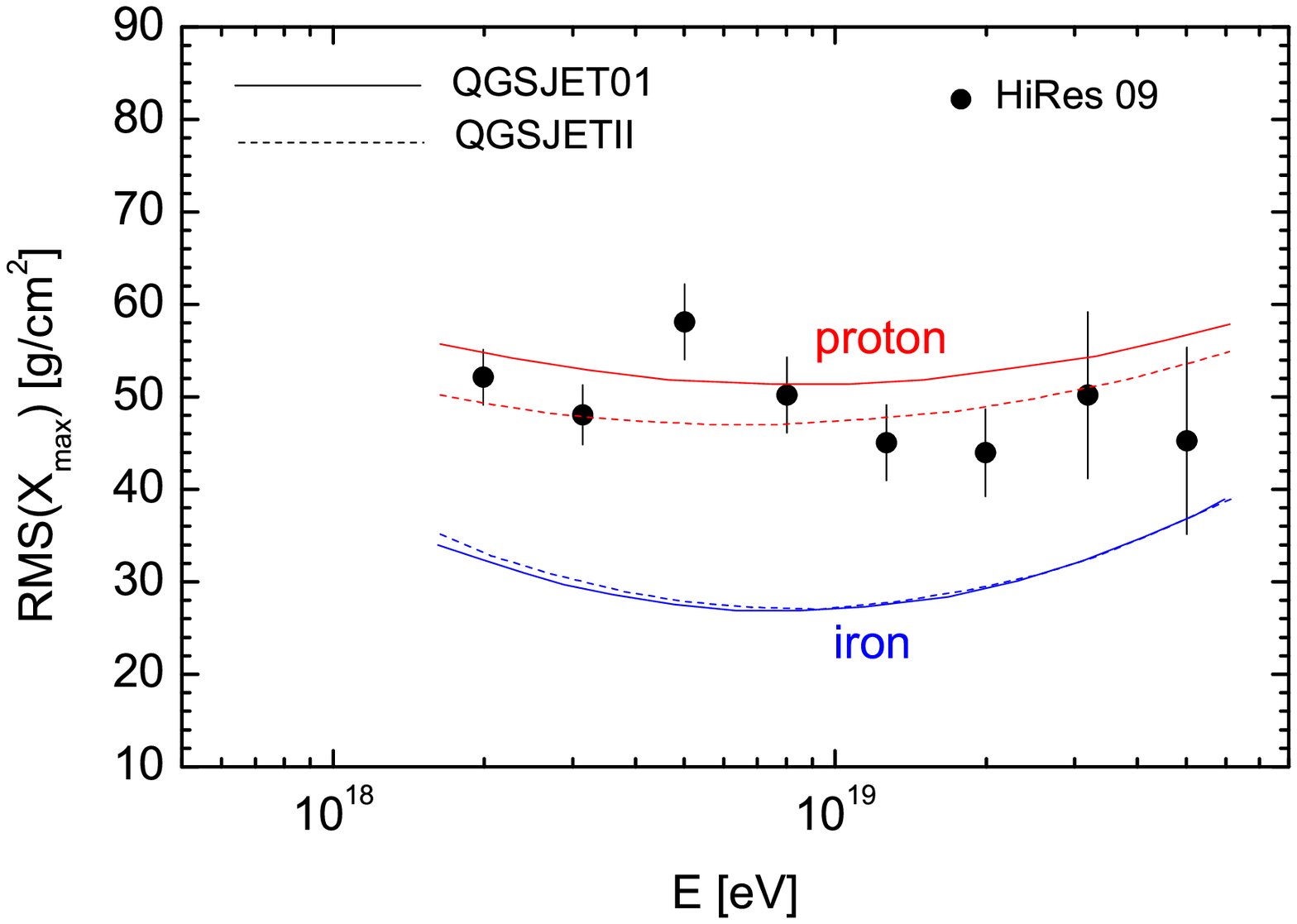}
 \end{minipage}
\newline \noindent
\medskip \hspace{-8mm}
 \begin{minipage}[ht]{66mm}\vspace{2mm}
 \includegraphics[width=66mm,height=45mm]{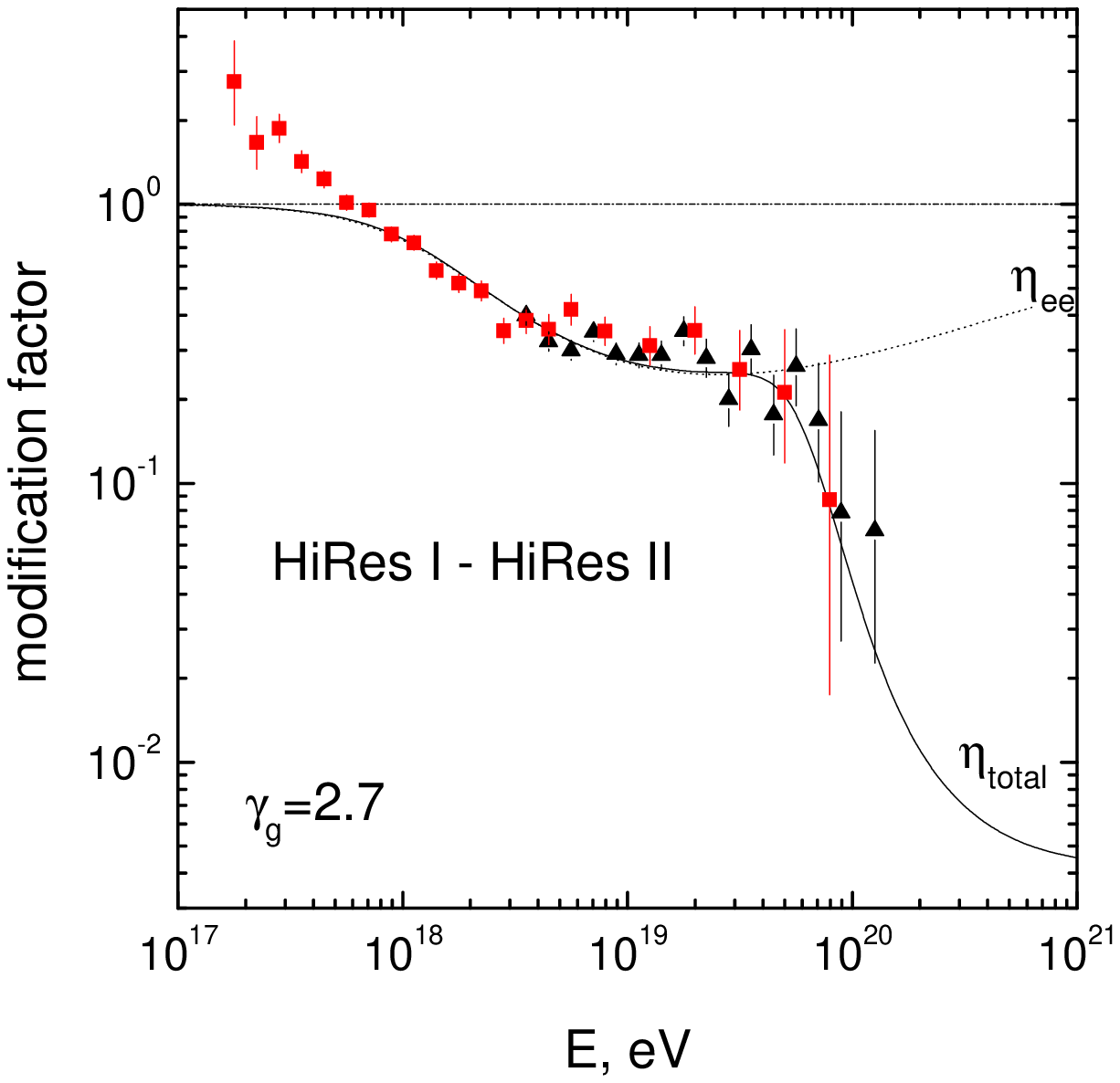}
 \end{minipage}
 \hspace{3mm}
 \begin{minipage}[h]{60mm}\medskip
 \includegraphics[width=59mm,height=45mm]{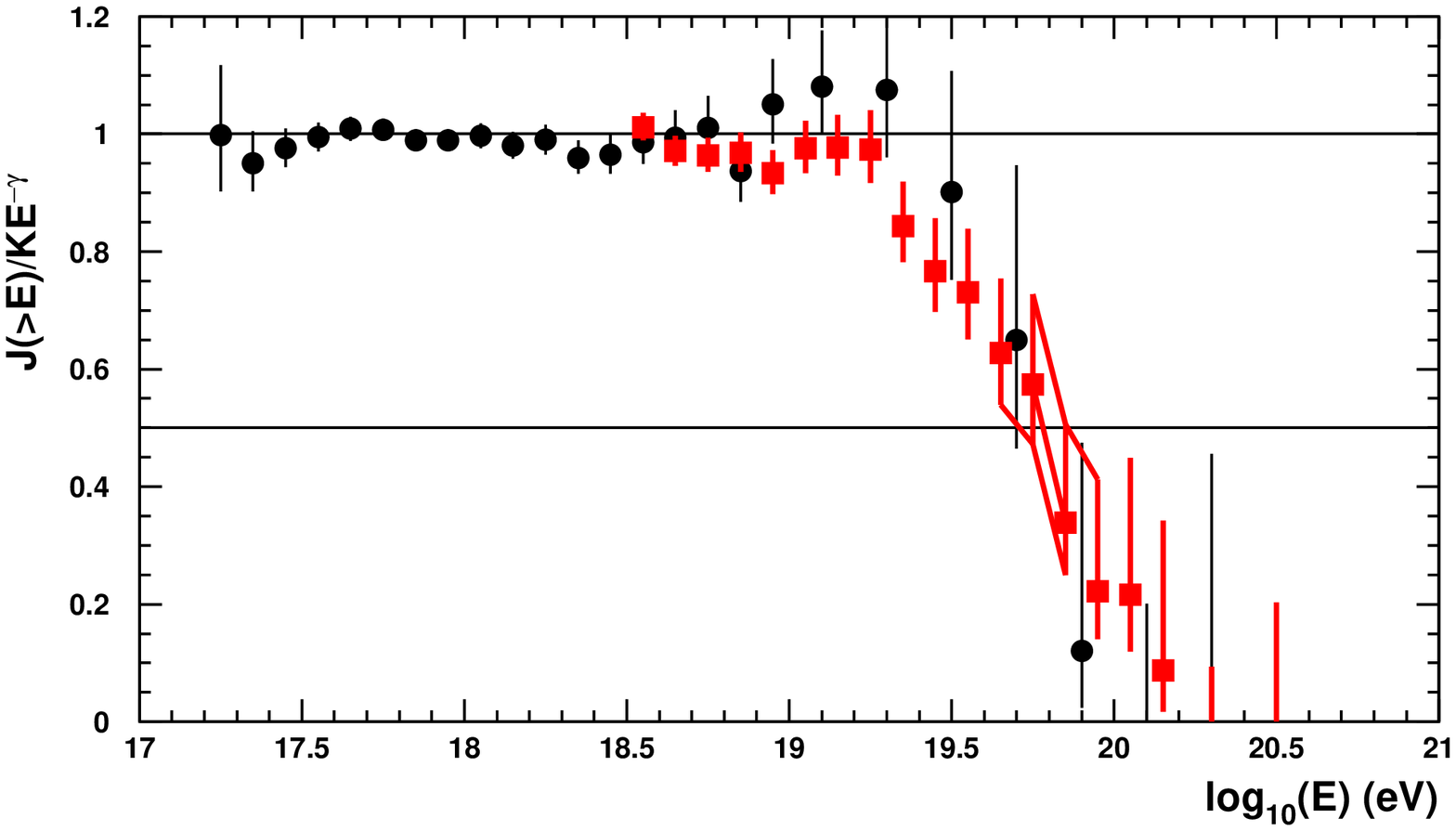}
\end{minipage}
 \vspace{-2 mm}%
\caption{ Mass composition and GZK cutoff as measured by the HiRes
detector. In two upper panels $\left\langle X_{\max}\right\rangle$
(left) and RMS (right) are presented as function of the energy.
Both agree with a pure proton composition, shown by curves labeled
'proton'. The left-lower panel shows differential energy spectrum
in terms of the modification factor. One can see a good agreement
with the predicted shape of the GZK cutoff. The right-lower panel
shows the quantity $E_{1/2}$ in the integral spectrum. This
energy, a characteristic of the GZK cutoff, is found as $E_{1/2}=
10^{19.73\pm 0.07}$~eV in good agreement with theoretical
prediction $E_{1/2}= 10^{19.72}$~eV (see the text).
} %
\label{fig7}
\end{center}
\end{figure} %
In this subsection we analyze data of the HiRes which provide a
strong evidence in favour of the GZK cutoff. These data are
supported also by the TA data \cite{GZK-TA}. The data of PAO will
be considered in the next subsection.

To interpret convincingly the spectrum steepening as the GZK cutoff
one must prove that ({\em i}) energy scale of the cutoff and its shape
correspond to theoretical predictions and ({\em ii}) the measured
mass composition is strongly dominated by protons. In HiRes the mass
composition is determined from $\left\langle X_{\rm max}\right\rangle(E)$,
average depth of atmosphere in $g/cm^2$, where a shower with energy $E$
reaches maximum, and RMS($X_{\max})$, which is the width of the distribution
over $X_{\max}$. These values measured by HiRes are displayed in
Fig.~\ref{fig7}. From the left-upper panel of Fig.~\ref{fig7}
one can see that the chemical composition changes from very heavy elements,
probably Iron, at $E \sim 0.1$~EeV, (data of HiRes-MIA \cite{hires-mia}) to
protons at $E\sim 1$~EeV (data of HiRes \cite{ankle-hires}). RMS$(X_{\max})$,
a very sensitive tool for mass composition, also provides evidence for a
proton-dominated composition at $E \gsim 1$~EeV and up to the highest
energies (see upper-right panel of Fig.~\ref{fig7}). Differential
energy spectrum of the GZK feature in the form of modification factor
(left-lower panel) is in a reasonably good agreement with the theoretical
prediction, though better statistics at higher energies is still needed for
a final conclusion.

The {\em integral energy spectrum} of UHE protons, $J_p(>E)$, has
another specific characteristic of the GZK cutoff, the energy $E_{1/2}$
\cite{BG88}. It is based on the observation that the calculated
integral spectrum below $50$~EeV is well approximated by a power-law
function: $J_p(>E) \propto E^{-\tilde{\gamma}}$. At high energy this
spectrum is steepening due to the GZK effect. The energy where this steep
part of the spectrum equals to the half of its power-law extrapolation,
$J_p(>E)=K E^{-\tilde{\gamma}}$, defines the value of $E_{1/2}$. This
quantity is found to be practically model-independent; it equals to
$E_{1/2} = 10^{19.72}\mbox{eV} \approx 52.5 $ EeV \cite{BG88}.
Fig.~\ref{fig7} demonstrates how the HiRes collaboration found
$E_{1/2}$ from observational data \cite{E_1/2hires}. The ratio of the
measured integral spectrum $J(>E)$ and the low-energy power-law
approximation $KE^{-\tilde{\gamma}}$ was plotted as a function of
energy. This ratio is practically constant in the energy interval $0.3 -
40$~EeV, indicating that the power-law approximation is a good fit,
indeed. At higher energy the ratio falls down and intersects the
horizontal line $0.5$ at the energy defined as $E_{1/2}$. It results in
$E_{1/2}= 10^{19.73\pm 0.07}$~eV, in an excellent agreement with the
predicted value.

Thus, one may conclude that the HiRes data presented in
Fig.~\ref{fig7} indicate the proton-dominated chemical
composition and the presence of the GZK cutoff in both differential and
integral spectra. The conclusion about proton composition is further
supported by the recent TA data \cite{GZK-TA}.
\begin{figure}\medskip
\begin{center}
 \begin{minipage}[h]{60mm}
 \includegraphics[width=60mm,height=40mm]{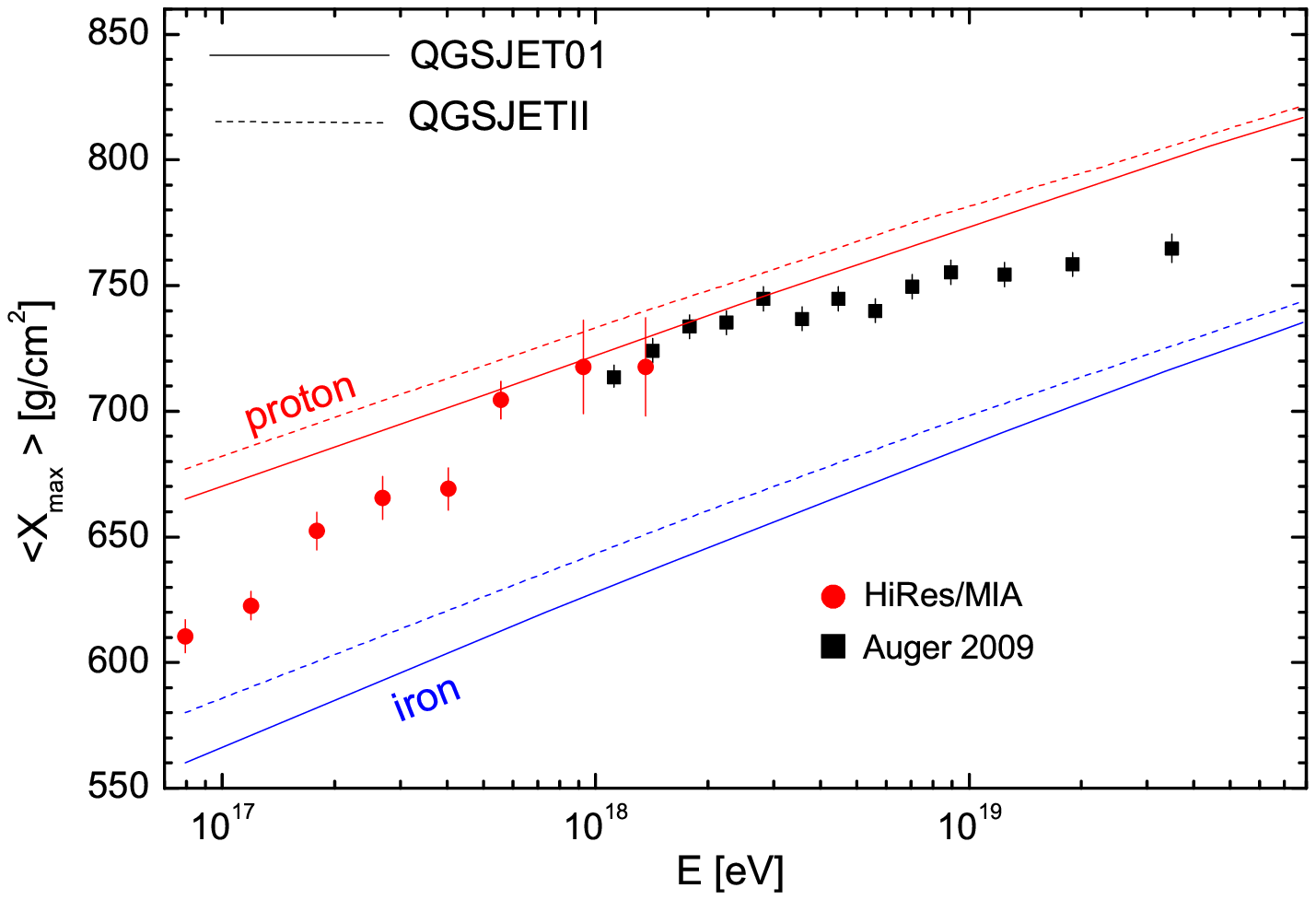}
 \end{minipage}
 \hspace{4mm}
 \begin{minipage}[h]{62mm}\vspace{1mm}\hspace{2mm}
 \includegraphics[width=61mm,height=43mm]{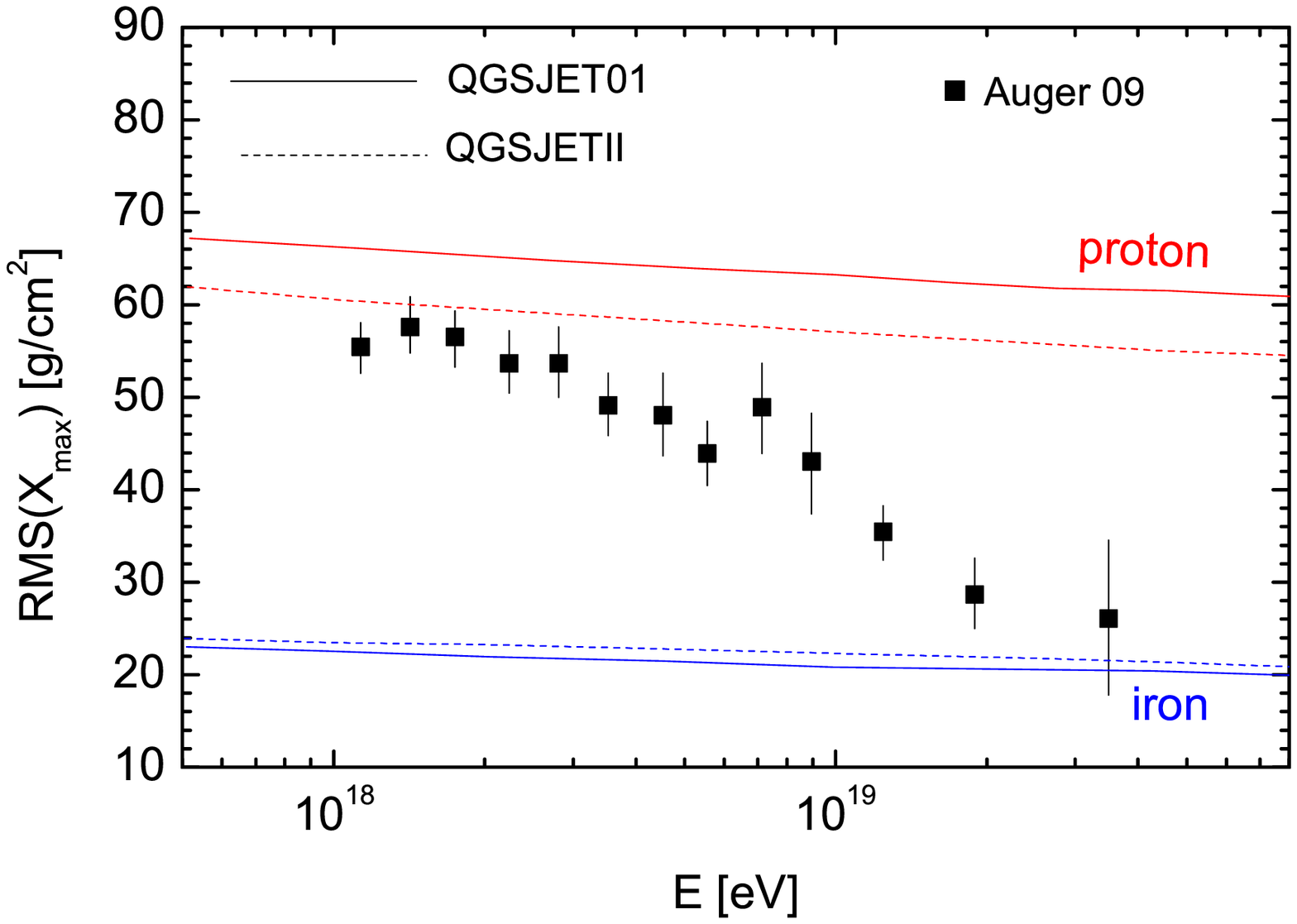}
 \end{minipage}
 \vspace{-2mm}
\caption{{\em Left panel:} Auger data \cite{auger-mass} for
$\left\langle X_{\max}\right\rangle$ as function of the energy
(left panel) and for RMS($X_{\max}$), the width of the
distribution over $X_{\max}$, (right panel). The calculated values
for protons and Iron are given according to QGSJET01 \cite{QGSJET}
and QGSJET II \cite{QGSJETII} models. One can see from the right
panel that RMS distribution becomes more narrow with increasing
energy which implies a progressively heavier mass composition. }
\label{fig8}
\end{center}
\end{figure}
\subsection{PAO data: energy spectrum and mass composition}
\label{sec:auger}
In subsection~\ref{energy-calibrator} we demonstrated that the
dip shape, as observed by PAO, can agree after recalibration with
energy spectra of HiRes/TA and other detectors (see
Figs.~\ref{fig5} and~\ref{fig6} ). The agreement
of the PAO and HiRes/TA  spectra is related to low energy part of the
energy spectrum in the right panel of Fig.~\ref{fig5}. At 
higher energies statistical uncertainties are too large to distinguish 
between the spectra.

While the HiRes and TA spectra are compatible with the GZK cutoff, the
Auger spectrum is not. The steepening in the PAO
spectrum  starts at energy $E=25.7 \pm 1.2$ EeV
\cite{Kampert,GZK-PAO2011}, lower than $E_{\rm GZK}
\simeq 50$ EeV, and in three successive energy bins in the interval
$35 - 52$~EeV the PAO flux is significantly lower than one predicted
for the GZK shape as shown in the right panel of Fig.~\ref{fig5}.
We could not reconcile the PAO cutoff shape with the GZK behavior by
including in the calculations different generation indices $\gamma_g$,
evolution regimes, low acceleration maximum energy $E_{\max}$, local
overdensity of sources etc.

This disagreement is quite natural for the PAO mass composition
which, in contrast to HiRes and TA, show strong dominance of
nuclei (see Fig.~\ref{fig8}). A steepening in the end of the
of the nuclei spectrum, as calculations show, is quite different
from that of protons (GZK cutoff). The most reliable data on mass
composition are given by elongation curve $X_{\max}(E)$ and
especially by RMS($X_{\max}(E))$. In the Auger data the latter
steadily decreases with energy and approaches the Iron value at $E
\approx 35$~EeV. Low RMS, i.e.\ small fluctuations, is a typical
and reliable feature of the heavy nuclei composition. These data
are further strengthened by other PAO measurements provided by
surface detectors. They allow to extract two other
mass-composition dependent quantities: the atmospheric depth
$\left\langle X_{\max}^{\mu}\right\rangle$, where muon-production
rate reaches maximum, and maximum zenith angle $\theta_{\max}$
determined by the signal rise-time in surface Cerenkov detectors.
Measurements of both quantities, \cite{Garcia-GamezICRC} and 
\cite{Gazon2012}, confirm the heavy mass composition
and its dependence on energy obtained with the help of
$\left\langle X_{\max}\right\rangle(E)$ and RMS($X_{\max}$),
 
Our further analysis of the Auger spectrum and mass composition is
based on the following two observations:\\*[1mm]
{\em (i)} According to the HiRes (Fig.~\ref{fig7}) and PAO
(Fig.~\ref{fig8}) data, the observed primaries at energy
$(1 - 3)$~EeV are predominantly protons or nuclei not heavier than
Helium.\\*[1mm]
These particles cannot be galactic, otherwise, as MC simulations
\cite{Giacinti} show, galactic anisotropy would be too large.\\*[1mm]
{\em (ii)}  The particles at higher energies are extragalactic nuclei
with the charge number $Z$ increasing with energy. \\*[1mm]
This observation is naturally explained by rigidity dependent
acceleration in the sources $E_i^{\max}=Z_i \times E_p^{\max}$,
since at each energy $E=ZE_p^{\max}$ the contribution of nuclei
with smaller $Z' < Z$ vanishes. It was demonstrated in \cite{disapp} 
that to avoid a proton dominance at
the highest energies, one must assume that the maximum energy of the 
accelerated protons is limited, $E_p^{\max} \lsim (4 - 10)$~EeV and
thus $E_{\rm Fe}^{\max} \sim 200$~EeV. The calculated spectrum \cite{disapp} 
explains well the PAO energy spectrum, however a problem remains to explain 
simultaneously energy spectrum and mass composition. Somewhat similar
calculations are performed in \cite{Allard2011} (see Fig.~4 there),  
and \cite{Taylor2011}.

%
%
\section{Status of ankle as transition from galactic to extragalactic 
cosmic rays.}
\label{transition-at-ankle}
As was discussed in section \ref{introduction} there are two features in the
observed CR spectrum, where transition from galactic to extragalactic 
cosmic rays is expected: the second knee and ankle. 

The {\em second knee} at energy $E_{\rm scn} \approx (0.4 - 0.7)$~EeV 
as transition feature corresponds well to the Standard Model (SM) of 
Galactic CRs, where maximum energy of acceleration corresponds to 
Iron knee, $E_{\max} \approx (0.08 - 0.1)$~EeV. 

The {\em ankle} is located at $E_a \approx (4 - 5)$~EeV according to 
all data, and as a feature of transition it needs another high-energy 
galactic component beyond SM. On the other hand the ankle appears 
automatically as the {\em dip feature} in Akeno-AGASA, HiRes, Yakutsk 
and TA data (see Fig.~\ref{fig4}) and also in  Auger data (Fig.\ref{fig5} 
right panel) after recalibration of energies with $\lambda=1.22$ and 
inclusion of evolution of the sources in theoretical prediction. At this 
comparison the power-law spectra approximation is not needed. Note that 
ankle in this case is a feature produced by interaction of extragalactic
protons with CMB. We will demonstrate below that interpretation of the
ankle as transition feature contradicts the observational data of the Auger, 
HiRes and TA  detectors.

The key data used in a proof is given by energy interval (1 - 3)~EeV 
where the observed mass composition in HiRes, TA and Auger agrees with 
pure proton composition and cannot be heavier than Helium (for Auger
data see Fig.~\ref{fig8}). In the ankle model of transition this interval  
is below the ankle at  $E_a \approx (4 - 5)$~EeV and thus must have 
galactic origin. Meanwhile, the Monte-Carlo simulations (see e.g. 
\cite{BGM2001}) and most convincingly the recent one \cite{Giacinti},
show that predicted anisotropy exclude the galactic protons and light 
nuclei at $E \gsim 1$~EeV as observed primary particles. This
conclusion is further strengthened by recent upper limit on the 
anisotropy observed by the Auger collaboration \cite{anisotropy-Auger}. 

In fact, the conclusion about ruling out the ankle as transition feature 
has been already made in \cite{ABBO} and more recently in \cite{ABG-rev} 
from analysis of elongation curve $X_{\max}(E)$. One can see from 
right-upper and right-lower panels of Fig.~10 in \cite{ABG-rev}, the 
strong contradiction of the ankle model with the $X_{\max}(E)$ data of 
Auger and HiRes in (1 - 5)~EeV energy interval. 

The analysis above proves that transition from galactic to extragalactic 
CRs occurs below 1~EeV, where the only visible spectral feature is 
the second knee. Together with the proton mass composition at (1 - 3)~EeV
observed in HiRes, TA and Auger experiments it gives one more support to 
the pair-production nature of the observed dip and the dip model developed
in \cite{BGGPL,BGGprd,Aletal}. 

An alternative explanation of the ankle observed in PAO at 
$E_a \approx 4$~EeV is given by transition at this energy from
extragalactic protons to extragalactic nuclei. However according to 
PAO observations \cite{auger-mass} the energy of this transition 
using the change in elongation rate and in RMS(E) is given by 
$E_{\rm tr}=2.40^{+0.42}_{-0.72}$~EeV to be compared with position of 
ankle $E_a=(4.17 \pm 0.1)$~EeV. If to assume $E_{\rm tr} \approx E_a$ 
the Auger spectrum and mass composition can be explained by 
disappointing model \cite{disapp}.  The transition from galactic 
CRs (Iron) to extragalactic protons occurs at the second knee.    
Transition from extragalactic protons to extragalactic nuclei occurs 
near the maximum of acceleration energy for protons 
$E_p^{\max} \approx 4$~EeV. This energy provides transition from
protons to heavier nuclei because their maximum energy 
$E_A^{\max}=Z E_p^{\max}$ is higher. The observed highest energy 
steepening can be explained as nuclei-photodisintegration cutoff.    
\section{Conclusions.}
\label{conclusions}
The most important result of analysis performed in this paper is the 
conclusion on  interpretation of {\em ankle}, observed at 
$E_a \sim 4$~EeV in all experiments. This feature is usually
interpreted as transition from galactic to extragalactic CRs. 
The presented analysis is based on consensus about mass composition 
measured by HiRes, TA and Auger detectors, which show that at energies 
(1 - 3)~EeV (i.e. below the ankle) the primaries are protons or nuclei 
not heavier than Helium. Having galactic origin, these particles must 
show the big anisotropy \cite{Giacinti} in contrast to low anisotropy 
measured recently \cite{anisotropy-Auger} by Auger detector. In fact 
this analysis only enhances the similar conclusion reached in papers
\cite{ABBO} and \cite{ABG-rev} about ruling out the models with 
transition at the ankle. using the measured elongation curves 
$X_{\max}(E)$.

In the light of these results, the transition must occur at lower
energies, most probably at the {\em second knee} from Iron galactic 
component to proton extragalactic component as the {\em dip} model 
predicts \cite{BGGPL,BGGprd,Aletal}. It strengthens further the dip 
model.   

The dip model is based on the dip as a signature of $e^+e^-$ 
pair-production in interaction of UHE protons with CMB: 
$p+\gamma_{\rm cmb} \to p+e^-+e^+$.  The characteristic which 
is sensitive mostly to interaction is given by {\em modification 
factor} $\eta(E)=J_p(E)/J_{\rm unm}(E)$, where unmodified proton flux 
$J_{\rm unm}(E)$ is calculated only with adiabatic energy losses taken 
into account, while $J_p(E)$ is calculated accounting for all energy
losses. One may see that interactions enter only numerator and thus
modification factor is very sensitive to interactions, while most 
other phenomena enter both, numerator and denominator, and thus they 
are suppressed or even cancelled in modification factor. The theoretical 
modification factor is presented in Fig.~\ref{fig3} and one can see
that it practically does not depend on the generation index $\gamma_g$.   
Another property of modification factor is $\eta(E) \to 1$ at 
$E \to 1$~EeV. Comparing the theoretical and experimental modification 
factors (see Fig.~\ref{fig4}) one can use only one free physical parameter  
for description about 25 bins in each of four experiments, however
agreement is quite good. It confirms that observed dip is really
produced by $p+\gamma_{\rm cmb} \to p+e^-+e^+$. 

One naturally expects that the calculated interaction signature in terms 
of modification factor cannot have the agreement with high-statistics 
observations with very good $\chi^2$, because the observational data include
the model-dependent features described by many parameters, such as
cosmological evolution of the sources, source separation etc. When
statistics increases, agreement with modification factor must become 
worse as it happened with Auger data. As the next step we performed the
{\em model-dependent} calculations which necessarily include many free
parameters describing all important physical effects. This analysis should
be done in terms of natural $E^3J(E)$ spectrum, where the model-dependent
features are not suppressed.   

The dip has the fixed energy position and shape, therefore it can serve 
as natural energy calibrator. Shifting the energy scale in each four 
experiments by factor $\lambda$, i.e. shifting each energy bin in each 
experiment $a$ by factor $\lambda_a$, one can reach $\chi^2_{\min}$ for 
each experiment. Interestingly enough that as a result of this procedure 
the Auger data show the dip (see right panel of Fig.~\ref{fig5}) and 
fluxes of all experiments coincide, as one can see in Fig.~\ref{fig6} and 
in Fig.~11 of Ref.~\cite{BGGprd}. We assume there $\lambda=1$ for HiRes 
experiment, because it has the correct scales in the form of beginning 
of GZK cutoff and $E_{1/2}$. 

The dip is valid for proton-dominated spectrum. Mixing of protons with 
$15\%$ of nuclei modifies the dip shape. 

The different mass composition presented as a result of direct
measurements in the HiRes and TA experiments, on one side, and in 
Auger experiments, on the other side, is at present the main conflict 
in UHECR. While the data of HiRes/TA show the proton-dominated mass
composition at all energies above 1~EeV, a few different methods of 
mass-composition measurement in Auger experiments demonstrate the 
nuclei-dominated composition starting from energy $E \sim 4$~EeV and
being steadily heavier as energy increases. Data of all three
experiments, HiRes, TA and Auger, coincide in narrow energy interval 
(1 - 3)~EeV, where composition is proton-dominated. Each of two 
possibilities can be viable within two contradicting models. 

The model for description of HiRes and TA data is a proton-dominated 
one with transition from galactic (Iron nuclei) to extragalactic 
(proton) component at the second knee. The presence of two features, 
the pair-production dip and GZK cutoff, are compulsory for such 
mass compositions, and these features are observed indeed as the shape of 
the dip and beginning of the GZK cutoff, together with value of $E_{1/2}$.
The dip model describes all these features numerically.

The Auger mass composition implies quite different scenario. The
transition from galactic to extragalactic CRs occurs at the second
knee ($E_{\rm skn} \approx (0.4 - 0.7)$~EeV) as the only visible 
spectral feature between the Iron knee (galactic feature) and 
the ankle (extragalactic feature). The ankle at $E_a \sim 4$~EeV 
may be produced by transition from extragalactic proton component 
at $E < 4$~EeV  to extragalactic heavy nuclei component, though
according to Auger data \cite{auger-mass} this transition occurs 
at lower energy $E \approx 2.4$~EeV. Proton component disappears 
because of low $E_p^{\max} \sim 4$~EeV at acceleration \cite{disapp}, 
and heavier nuclei progressively dominate at higher energies because 
of the higher acceleration energies $E_A^{\max}= Z E_p^{\max}$, or
because of diffusion, or both. The highest-energy spectrum suppression,
observed by Auger at $E_{\rm supp}=25.7 \pm 1.2$~EeV \cite{GZK-PAO2011}  
is too low for GZK cutoff. It can be explained as
nuclei-photodisintegration cutoff.  

The Auger scenario is characterised by very specific behaviour 
of $X_{\max}(E)$ and RMS(E) curves since protons appear in a
relatively small energy range (1 - 3)~EeV. In this range $X_{\max}$
and RMS reach the highest values, typical for protons, with lower values 
starting at both ends of these interval, because of transition to 
galactic Iron (lower end) and extragalactic nuclei (upper end). 
Intersection of galactic spectrum (Iron) with extragalactic protons 
corresponds to maximum of RMS. 

All three biggest experiments use the fluorescent detectors as the
main tool of measurements. The contradicting results are connected with
some difference in the data analysis. Close collaboration between Telescope 
Array and Auger teams may resolve this problem in the nearest future. 
The role of new experiments is less significant: already now there two 
experiments which confirm proton-dominated composition and one -
nuclei-dominated composition. Science is a choice, but not elections.  
\section*{Acknowledgment}
This paper is based on the joint works and many discussions with
my co-authors Roberto Aloisio, Askhat Gazizov and Svetlana Grigorieva.
I am mostly grateful to them for efficient and pleasant collaboration.
The work was partly supported by Ministry of Science and
Education of Russian Federation (agreement  8525).

\section*{References}

\end{document}